\documentclass[11pt]{article}
\usepackage{times}
\usepackage[dvips]{graphicx}

\def	\ddscat	{{\bf DDSCAT.6.0}}
\begin{document}
\title{\vspace*{-3em} {\normalsize
This document can be cited as: Draine, B.T., and Flatau, P.J. 2003, \\
``User Guide for the Discrete Dipole Approximation Code DDSCAT.6.0'', \\
\vspace*{-0.6em} http://arxiv.org/abs/astro-ph/0309069} \\ \vspace*{2em}
	{\bf User Guide for the Discrete Dipole} \\
	{\bf Approximation Code \ddscat}}
                                
\author{Bruce T. Draine \\
	Princeton University Observatory \\
	Princeton NJ 08544-1001 \\
	({\tt draine@astro.princeton.edu})
	\\
	and \\
	\\
	Piotr J. Flatau \\
        University of California, San Diego \\
	Scripps Institution of Oceanography \\
	La Jolla CA 92093-0221 \\
	({\tt pflatau@ucsd.edu})
	}
\date{last revised: 2003 September 2}
\maketitle
\abstract{
	\ddscat\ is a freely available 
	software package which applies the
	``discrete dipole approximation'' (DDA) to calculate scattering
	and absorption of electromagnetic waves by targets with arbitrary
	geometries and complex refractive index.  The DDA approximates
	the target by an array of polarizable points.
	\ddscat\ allows
	accurate calculations of electromagnetic scattering from targets
	with ``size parameters'' $2\pi a/\lambda < 15$ provided the
	refractive index $m$ is not large compared to unity ($|m-1| < 1$).
	\ddscat\ includes the option of using the {\tt FFTW}
	(Fastest Fourier Transform in the West) package.
	\ddscat\ also includes support for
	{\tt MPI} (Message Passing Interface),
	permitting parallel calculations on multiprocessor systems.

	The {{\bf DDSCAT}} package is written in Fortran 
	and is highly portable.
	The program supports calculations for a variety of target geometries
	(e.g., ellipsoids, regular tetrahedra, rectangular solids, 
	finite cylinders, hexagonal prisms, etc.).
	Target materials may be both inhomogeneous and anisotropic.
	It is straightforward for the user to ``import'' arbitrary
	target geometries into the code, and relatively straightforward
	to add new target generation capability to the package.
	{{\bf DDSCAT}} automatically calculates total cross sections
	for absorption and scattering and selected elements of the
	Mueller scattering intensity 
	matrix for
	specified orientation of the target relative to the incident wave,
	and for specified scattering directions.

	This User Guide explains how to use \ddscat\ to carry
	out electromagnetic scattering calculations.  
	CPU and memory requirements are described.
	}

\tableofcontents
\newpage
\section{Introduction\label{intro}}

\ddscat\ is a Fortran software package to calculate scattering and
absorption of electromagnetic waves by targets with arbitrary geometries
using the ``discrete dipole approximation'' (DDA).
In this approximation the target is replaced by an array of point dipoles
(or, more precisely, polarizable points); the electromagnetic scattering
problem for an incident periodic wave interacting with this array of
point dipoles is then solved essentially exactly.
The DDA (sometimes referred to as the ``coupled dipole approximation'') 
was apparently first proposed by Purcell \& Pennypacker (1973).
DDA theory was reviewed and developed further by Draine (1988), 
Draine \& Goodman (1993), and recently reviewed by Draine \& Flatau (1994)
and Draine (2000).

\ddscat\ is a Fortran implementation of the DDA 
developed by the authors.
It is intended to be a versatile tool, suitable for a wide variety
of applications ranging from interstellar dust to atmospheric aerosols.
As provided, \ddscat\ should be usable 
for many applications without
modification, but the program is written in a modular form, so that
modifications, if required, should be fairly straightforward.

The authors make this code openly available to others, in the hope that it
will prove a useful tool.  We ask only that:
\begin{itemize}
\item If you publish results obtained using {{\bf DDSCAT}}, please consider 
	acknowledging the source of the code.

\item If you discover any errors in the code or documentation, 
	please promptly communicate them to the authors.

\item You comply with the ``copyleft" agreement (more formally, the 
	GNU General Public License) of the Free Software Foundation: you may 
	copy, distribute, and/or modify the software identified as coming 
	under this agreement. 
	If you distribute copies of this software, you must give the 
	recipients all the rights which you have. 
	See the file {\tt doc/copyleft} distributed with the DDSCAT software.
\end{itemize}
We also strongly encourage you to send email to the authors identifying  
yourself as a user of DDSCAT;  this will enable the authors to notify you of
any bugs, corrections, or improvements in DDSCAT.

The current version, \ddscat, 
uses the DDA formulae from Draine (1988), with
dipole polarizabilities determined from the Lattice Dispersion
Relation (Draine \& Goodman 1993). 
The code incorporates Fast Fourier Transform (FFT) methods
(Goodman, Draine, \& Flatau 1991).

This User Guide assumes that you have already obtained the Fortran
source code for \ddscat\ either by download from
{\tt http://www.astro.princeton.edu/$\sim$draine} or by
following the instructions 
in the {\tt README} file.\footnote{
	To obtain the {\tt README} file: \newline
	(1) anonymous {\tt ftp} to {\tt astro.princeton.edu} ,\newline
	(2) {\tt cd draine/scat/DDA/ver6}, and\newline
	(3) {\tt get README}.
	}
We refer you to the list of references at the end of this document for 
discussions
of the theory and accuracy of the DDA [first see the
recent reviews by Draine and Flatau (1994) and Draine (2000)]. 
In \S\ref{sec:whats_new} we 
describe the principal changes between \ddscat\ and the previous 
releases.\footnote{
	The previous ``official releases'' were 
	\begin{itemize}
	\vspace*{-0.4em}
	\item{\bf DDSCAT.4b},
	\vspace*{-0.4em} 
	\item{\bf DDSCAT.4c}  --while never announced, {\tt DDSCAT.4c} 
		was made available to a number of interested users.
	\vspace*{-0.4em}
	\item{\bf DDSCAT.5a7}
	\vspace*{-0.4em}
	\item{\bf DDSCAT.5a8}
	\vspace*{-0.4em}
	\item{\bf DDSCAT.5a9}
	\vspace*{-0.4em}
	\item{\bf DDSCAT.5a10}
	\end{itemize}
	}
The succeeding sections contain instructions for:
\begin{itemize}
\item compiling and linking the code;
\item running a sample calculation;
\item understanding the output from the sample calculation;
\item modifying the parameter file to do your desired calculations;
\item specifying target orientation;
\item changing the {\tt DIMENSION}ing of the source code to accommodate 
your desired calculations.
\end{itemize}
The instructions for compiling, linking, and running will be appropriate for a
UNIX system; slight changes will be necessary for non-UNIX sites, but they are
quite minor and should present no difficulty.

Finally, the current version of this 
User Guide can be obtained from\newline
{\tt http://arxiv.org/abs/astro-ph/0008151} -- you will be offered
the options of downloading 
	\begin{itemize}
	\item 
	Latex source
	\item 
	Postscript
	\item 
	Other formats -- click on this to obtain the UserGuide as a PDF file.
	\end{itemize}

\section{Applicability of the DDA\label{sec:applicability}}
The principal advantage of the DDA is that it is completely flexible 
regarding the geometry of the target, being limited only by the need to 
use an interdipole separation $d$ small compared to 
(1) any structural lengths in the target, and
(2) the wavelength $\lambda$.
Numerical studies (Draine \& Goodman 1993; Draine \& Flatau 1994; Draine 2000)
indicate that the
second criterion is adequately satisfied if
\begin{equation}
|m|kd< 1~~~,
\label{eq:mkd_max}
\end{equation}
where $m$ is the complex refractive index of the target
material, and $k\equiv2\pi/\lambda$, where $\lambda$ is the wavelength
{\it in vacuo}.
However, if accurate calculations of the scattering phase function
(e.g., radar or lidar cross sections)
are desired,
a more conservative criterion 
\begin{equation}
|m|kd < 0.5
\end{equation}
will ensure that differential scattering cross sections
$dC_{\rm sca}/d\Omega$ are accurate to within a few percent of the
average differential scattering cross section $C_{\rm sca}/4\pi$
(see Draine 2000).

Let $V$ be the target volume.
If the target is represented by an array of $N$ dipoles, located on
a cubic lattice with lattice spacing $d$,
then 
\begin{equation}
V=Nd^3 ~~~.
\end{equation}
We characterize the size of the target by the ``effective radius''
\begin{equation}
a_{\rm eff}\equiv(3V/4\pi)^{1/3} ~~~,
\end{equation}
the radius of an equal volume sphere.
A given scattering problem is then characterized by the
dimensionless ``size parameter''
\begin{equation}
x\equiv ka_{\rm eff} = \frac{2\pi a_{\rm eff}}{\lambda} ~~~.
\end{equation}
The size parameter can be related to $N$ and $|m|kd$:
\begin{equation}
x\equiv{2\pi a_{\rm eff}\over\lambda} =
{62.04\over|m|}\left({N\over10^6}\right)^{1/3} \cdot |m|kd ~~~.
\end{equation}
Equivalently, the target size can be written
\begin{equation}
a_{\rm eff} = 9.873 {\lambda\over|m|}\left({N\over10^6}\right)^{1/3}
\cdot |m|kd~~~.
\end{equation}
Practical considerations of CPU speed and computer memory currently 
available on scientific workstations typically
limit the number
of dipoles employed to $N < 10^6$ (see \S\ref{sec:memory_requirements}
for limitations on $N$ due to available RAM); 
for a given $N$, the limitations on $|m|kd$ 
translate into limitations on the ratio of target size to wavelength.

\noindent
For calculations of total cross sections $C_{\rm abs}$ and $C_{\rm sca}$,
we require $|m|kd < 1$:
\begin{equation}
a_{\rm eff} < 9.88 {\lambda\over |m|}\left({N\over10^6}\right)^{1/3}
{\rm ~~or~~} x < {62.04\over|m|}\left({N\over10^6}\right)^{1/3} ~~~.
\end{equation}
For scattering phase function calculations, we require $|m|kd < 0.5$:
\begin{equation}
a_{\rm eff} < 4.94 {\lambda\over |m|}\left({N\over10^6}\right)^{1/3}
{\rm ~~or~~} x < {31.02\over|m|}\left({N\over10^6}\right)^{1/3} ~~~.
\end{equation}

It is therefore clear that the DDA is not suitable for very large values of
the size parameter 
$x$, or very large values of the refractive index $m$.
The primary utility of the DDA is for scattering by dielectric 
targets with sizes comparable to the wavelength.
As discussed by Draine \& Goodman (1993), Draine \& Flatau (1994), and
Draine (2000),
total cross sections calculated with the DDA are 
accurate to a few percent provided
$N>10^4$ dipoles are used, criterion (\ref{eq:mkd_max}) is satisfied,
and $|m-1|< 2$.

Examples illustrating the accuracy of the DDA are shown in 
Figs.\ \ref{fig:Qm=1.33+0.01i}--\ref{fig:Qm=2+i} which show overall
scattering and absorption efficiencies as a function of wavelength for
different discrete dipole approximations to a sphere, with $N$ ranging
from 304 to 59728.
The DDA calculations assumed radiation incident along the (1,1,1)
direction in the ``target frame''.
Figs.\ {\ref{fig:dQdom=1.33+0.01i}--\ref{fig:dQdom=2+i} show the scattering
properties calculated with the DDA for $x=ka=7$.
Additional examples can be found in Draine \& Flatau (1994) and Draine (2000).

\begin{figure}
\centerline{\includegraphics[width=8.3cm]{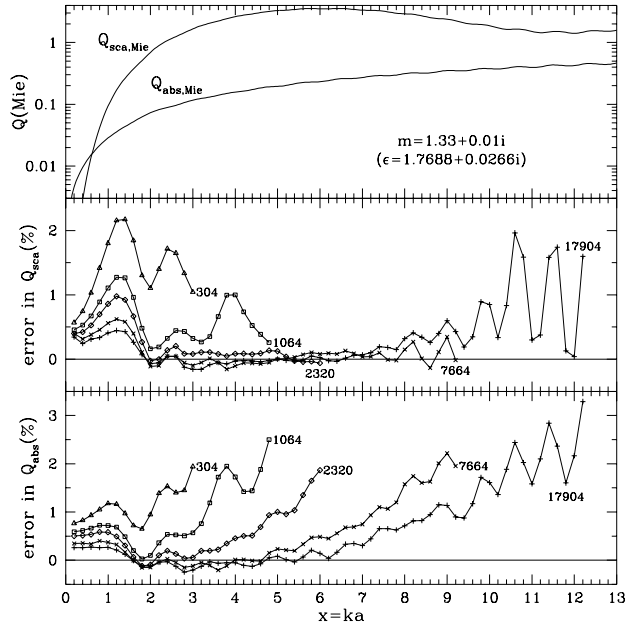}}
\caption{Scattering and absorption for a sphere with
	$m=1.33+0.01i$.  The upper panel shows the exact values of $Q_{\rm sca}$
	and $Q_{\rm abs}$, obtained with Mie theory, as functions of $x=ka$.
	The middle and lower panels show fractional errors in $Q_{\rm sca}$ and
	$Q_{\rm abs}$, obtained using {{\bf DDSCAT}}\ with polarizabilities obtained
	from the Lattice Dispersion Relation, and labelled by the number $N$
	of dipoles in each pseudosphere.
	After Fig.\ 1 of Draine \& Flatau (1994).}
	\label{fig:Qm=1.33+0.01i}
\end{figure}

\begin{figure}
\centerline{\includegraphics[width=8.3cm]{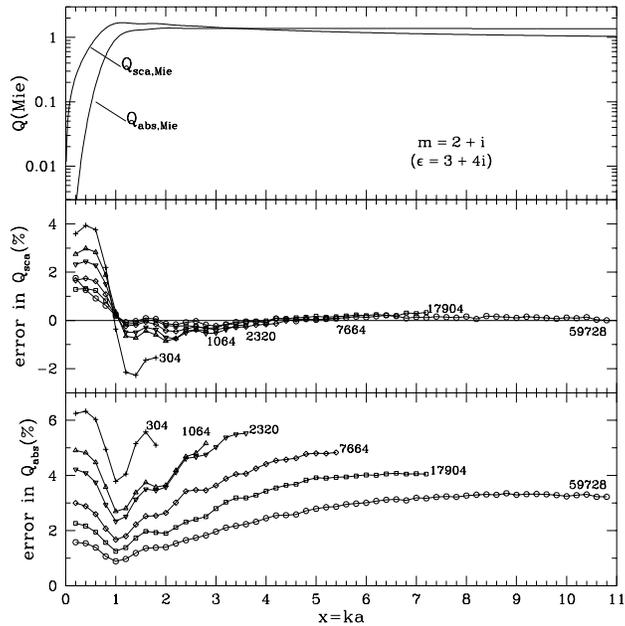}}
\caption{Same as Fig.\ \protect{\ref{fig:Qm=1.33+0.01i}},
	but for $m=2+i$. After Fig.\ 2 of Draine \& Flatau (1994).}
	\label{fig:Qm=2+i}
\end{figure}
\begin{figure}
\centerline{\includegraphics[width=8.3cm]{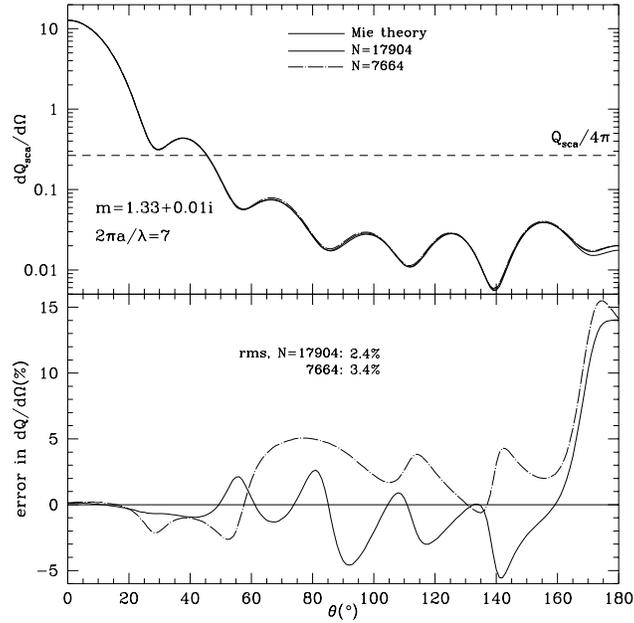}}
\caption{Differential scattering cross section for 
	$m=1.33+0.01i$ pseudosphere and $ka=7$.
	Lower panel shows fractional error compared to exact Mie theory
	result.
	The $N=17904$ pseudosphere has $|m|kd=0.57$, and an rms fractional
	error in $d\sigma/d\Omega$ of 2.4\%.
	After Fig.\ 5 of Draine \& Flatau (1994).}
	\label{fig:dQdom=1.33+0.01i}
\end{figure}
\begin{figure}
\centerline{\includegraphics[width=8.3cm]{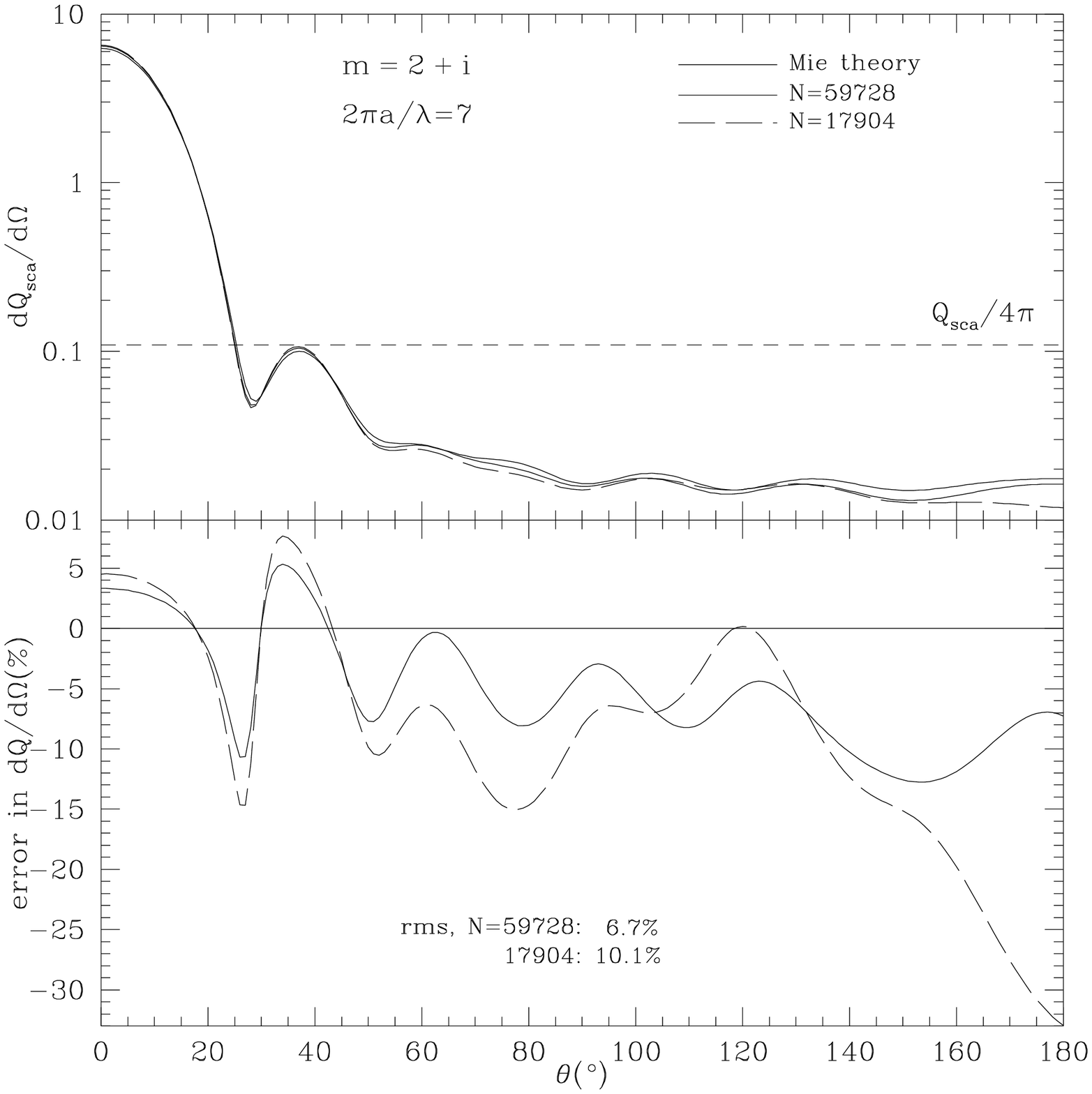}}
\caption{Same as Fig.\ \protect{\ref{fig:dQdom=1.33+0.01i}}
	but for $m=2+i$.
	The $N=59728$ pseudosphere has $|m|kd=0.65$, and an rms fractional
	error in $d\sigma/d\Omega$ of 6.7\%.
	After Fig.\ 8 of Draine \& Flatau (1994).}
	\label{fig:dQdom=2+i}
\end{figure}

\section{DDSCAT.6.0\label{sec:DDSCATvers}}
\subsection{What Does It Calculate?}
\ddscat, like previous versions of {{\bf DDSCAT}},
solves the problem of scattering and 
absorption by an array of
polarizable point dipoles interacting with a monochromatic plane wave incident
from infinity.  
\ddscat\ has the capability of automatically generating dipole
array representations for a variety of target geometries 
(see \S\ref{sec:target_generation}) and can also accept dipole
array representations of targets supplied by the user (although
the dipoles must be located on a cubic lattice).
The incident plane wave can have arbitrary elliptical
polarization (see \S\ref{sec:incident_polarization}), 
and the target can be arbitrarily oriented relative to the
incident radiation (see \S\ref{sec:target_orientation}).
The following quantities are calculated by \ddscat\ :
\begin{itemize}
\item Absorption efficiency factor $Q_{\rm abs}\equiv C_{\rm abs}/\pi a_{\rm eff}^2$,
where $C_{\rm abs}$ is the absorption cross section;
\item Scattering efficiency factor $Q_{\rm sca}\equiv C_{\rm sca}/\pi a_{\rm eff}^2$,
where $C_{\rm sca}$ is the scattering cross section;
\item Extinction efficiency factor $Q_{\rm ext}\equiv Q_{\rm sca}+Q_{\rm abs}$;
\item Phase lag efficiency factor $Q_{\rm pha}$, defined so that the phase-lag
(in radians) of a plane wave after propagating a distance $L$ is just
$n_{t}Q_{\rm pha}\pi a_{\rm eff}^2 L$, 
where $n_t$ is the number density of targets.
\item The 4$\times$4 Mueller scattering intensity matrix $S_{ij}$ 
describing the complete
scattering properties of the target for scattering directions specified
by the user.
\item Radiation force efficiency vector ${\bf Q}_{\rm rad}$ 
(see \S\ref{sec:torque_calculation}).
\item Radiation torque efficiency vector ${\bf Q}_\Gamma$
(see \S\ref{sec:torque_calculation}).
\end{itemize}
\subsection{Application to Targets in Dielectric Media}
	\label{sec:target_in_medium}}
Let $\omega$ be the angular frequency of the incident radiation.
For many applications, the target is essentially {\it in vacuo}, in which
case the dielectric function $\epsilon$ which the user should supply
to {\tt DDSCAT} is the actual complex dielectric function 
$\epsilon_{\rm target}(\omega)$,
or complex refractive index 
$m_{\rm target}(\omega)=\sqrt{\epsilon_{\rm target}}$ 
of the target material.

However, for many applications of interest (e.g., marine optics, or
biological optics) the ``target'' body is embedded in a (nonabsorbing)
dielectric medium,
with (real) dielectric function $\epsilon_{\rm medium}(\omega)$, or
(real) refractive index $m_{\rm medium}(\omega)=\sqrt{\epsilon_{\rm medium}}$.
{{\bf DDSCAT}} is fully applicable to these scattering problems, except
that:
\begin{itemize}
\item The ``dielectric function'' or ``refractive index'' 
supplied to {{\bf DDSCAT}} should be
the {\it relative} dielectric function
\begin{equation}
\epsilon(\omega) = 
\frac{\epsilon_{\rm target}(\omega)}{\epsilon_{\rm medium}(\omega)}
\end{equation}
or {\it relative} refractive index:
\begin{equation}
m(\omega) =
\frac{m_{\rm target}(\omega)}{m_{\rm medium}(\omega)} .
\end{equation}
\item The wavelength $\lambda$ specified in {\tt ddscat.par} should be the
wavelength {\it in the medium}:
\begin{equation}
\lambda = \frac{\lambda_{vac}}{m_{\rm medium}},
\label{eq:lambda_medium}
\end{equation}
where $\lambda_{vac}=2\pi c/\omega$ is the wavelength {\it in vacuo}.
\end{itemize}
The absorption, scattering, extinction, and phase lag 
efficiency factors $Q_{\rm abs}$,
$Q_{\rm sca}$, and $Q_{\rm ext}$ calculated by {{\bf DDSCAT}}
will then be equal to the physical
cross sections for absorption, scattering, and extinction divided by
$\pi a_{\rm eff}^2$ -- e.g., the attenuation coefficient for radiation
propagating through a medium with a density $n_{t}$ of
scatterers will be just
$\alpha = n_{t}Q_{\rm ext}\pi a_{\rm eff}^2$.
Similarly, the phase lag (in radians) after propagating a distance $L$ will be
$n_{t}Q_{\rm pha}\pi a_{\rm eff}^2$.

The elements $S_{ij}$ of the 4$\times$4 Mueller scattering matrix ${\bf S}$
calculated by {{\bf DDSCAT}}
will be correct for scattering in the medium:
\begin{equation}
{\bf I}_{\rm sca} = 
\left(\frac{\lambda}{2\pi r}\right)^2 
{\bf S}\cdot {\bf I}_{in} ,
\end{equation}
where ${\bf I}_{in}$ and ${\bf I}_{\rm sca}$ are the Stokes vectors for
the incident and scattered light (in the medium), 
$r$ is the distance from the target, and
$\lambda$ is the wavelength in the medium (eq.\ \ref{eq:lambda_medium}).
See \S\ref{sec:mueller_matrix} for a detailed discussion of the
Mueller scattering matrix.

The time-averaged 
radiative force and torque (see \S\ref{sec:torque_calculation}) on a
target in a dielectric medium are
\begin{equation}
{\bf F}_{\rm rad} = {\bf Q}_{pr}\pi a_{\rm eff}^2 u_{\rm rad} ~~~,
\end{equation}
\begin{equation}
{\bf \Gamma}_{\rm rad} = 
{\bf Q}_\Gamma \pi a_{\rm eff}^2 u_{\rm rad} \frac{\lambda}{2\pi} ~~~,
\end{equation}
where the time-averaged energy density is
\begin{equation} 
u_{\rm rad}=\epsilon_{\rm medium} \frac{E_0^2}{8\pi} ~~~ ,
\end{equation}
where $E_0\cos(\omega t+\phi)$
is the electric field of the incident plane wave in the medium.

\bigskip
\section{What's New?\label{sec:whats_new}}
\ddscat\ differs from
{{\bf DDSCAT.5a}}\ in the following ways:
\begin{enumerate}
\item Via the parameter file {\tt ddscat.par}, the user can now select up
	to 9 elements of the Muller scattering matrix $S_{ij}$ to be
	printed out.
\item The maximum number of iterations allowed has been increased from 300
	to 10000, since {\bf DDSCAT} is now increasingly employed
	for computations that converge relatively slowly: 
	large numbers of dipoles, large values of
	the scattering parameter, or large values of the refractive index.
	The number of iterations used for a solution is recorded in the
	{\tt ww}{\it aa}{\tt r}{\it bb}{\tt k}{\it ccc}{\tt .sca} output
	files.
\item A new target option is available: {\tt LYRSLB} (a
	rectangular slab with up to 4 different material layers).
\item User must now specify (in {\tt ddscat.par} 
	which {\tt IWAV}, {\tt IRAD}, {\tt IWAV} to
	start with.
	The ``default'' choice will be {\tt 0  0  0}, but when
	computations have been interupted by, e.g., power outage, there
	will be occasions when it is convenient to be able to resume with the
	first incomplete calculation.
\item A new FFT option is supported: {\tt FFTW21}.  This invokes the
	{\tt FFTW} (``Fastest Fourier Transform in the West'') package
	of Frigo and Johnson.  The calls are compatible with versions 2.1.x
	of FFTW.
	Instructions on compiling and linking to
	include {\tt FFTW} are given below (\S\ref{subsec:support_for_FFTW}).
\item FFT options {\tt BRENNR} and {\tt TMPRTN} have been eliminated
	as they seem to offer no advantages relative to {\tt GPFAFT}
	or {\tt FFTW21}.
\item {\tt MPI} is now supported.  Users can now carry out
	parallel calculations using \ddscat\ on multiprocessor
	systems in conformance with the {\tt MPI} 
	(Message Passing Interface) standards (\S\ref{sec:MPI}).
	This of course requires that {\tt MPI} be already installed on
	the system.
\end{enumerate}

\section{Obtaining the Source Code\label{sec:downloading}}
The easiest way to download the source code is from the DDSCAT home page:\\
\hspace*{2em}{\tt http://www.astro.princeton.edu/$\sim$draine/DDSCAT.html}\\
where you
can obtain the file {\tt ddscat6.0.tar.gz} -- a ``gzipped tarfile''
containing the complete source code and documentation.
This can also be obtained by anonymous {\tt ftp} from
{\tt astro.princeton.edu}, directory {\tt draine/scat/ddscat/ver6.0}.
Note that it is a compressed (binary) file.

The source code can be installed as follows.
First, place the file {\tt ddscat6.0.tar.gz}
in the directory where you would like {\bf DDSCAT}
to reside.  You should have at least 5 Mbytes of disk space available.

If you are on a Linux system, you should be able to type\\
\hspace*{2em}{\tt tar xvzf ddscat6.0.tar.gz}\\
which will ``gunzip'' the tarfile and then ``extract'' the files from the
tarfile.  If your version of ``tar'' doesn't support the ``z'' option
(e.g., you are running Solaris) then try\\
\hspace*{2em}{\tt zcat ddscat6.0.tar.gz | tar xvf -}\\
If neither of the above work on your system, try the two-stage procedure\\
\hspace*{2em}{\tt gunzip ddscat6.0.tar.gz}\\
\hspace*{2em}{\tt tar xvf ddscat6.0.tar}\\
(The disadvantage of the two-stage procedure is that it uses more disk
space, since after the second step you will have the uncompressed
tarfile {\tt ddscat6.0.tar} -- about 3.8 Mbytes --
in addition to all the files you have extracted
from the tarfile -- another 4.6 Mbytes).

Any of the above approaches should 
create subdirectories {\tt src, doc, misc,} and {\tt IDL}.
The source code will be in subdirectory {\tt src}, and documentation
in subdirectory {\tt doc}.

If you are running Windows on a PC, you will need the ``{\tt winzip}''
program, which can be downloaded from
{\tt http://www.winzip.com}.  {\tt winzip} should be able to ``unzip'' the
gzipped tarfile {\tt ddscat6.0.tar.gz} and ``extract'' the various files
from it automatically.

\section{Compiling and Linking\label{sec:compiling}}
In the discussion below, it will be assumed that the source 
code for \ddscat\ has been installed in a directory {\tt DDA/src}.
To compile the code on a Unix system, position yourself in the 
directory {\tt DDA/src}.  
The {\tt Makefile} supplied has compiler options set as appropriate
for use of the {\tt g77} compiler under the Linux operating system.
If you have a different Fortran compiler, you will probably need 
to edit {\tt Makefile} to provide the desired compiler options.
If you are using an operating system other than Linux, you may need
to change the Makefile choice of {\tt timeit}.
{\tt Makefile} contains
samples of compiler options for selected operating systems, including
Linux, 
HP AUX, IBM AIX, and SGI IRIX operating systems.

So far as we know, there are only two operating-system-dependent aspects of
\ddscat: (1) the device number to use for ``standard output", and 
(2) the {\tt TIMEIT} routine.
There are, in addition, three installation-dependent aspects to the code: 
\begin{itemize}
\item the
procedure for linking to the {\tt FFTW} library 
(see \S\ref{subsec:support_for_FFTW})
\item the procedure for linking to the {\tt netCDF} library 
(see \S\ref{subsec:netCDF} for
discussion of the possibility of writing binary files using the
machine-independent {\tt netCDF} standard).
\item the procedure for linking to the {\tt MPI} library
(see \S\ref{subsec:compiling_mpi}).
\end{itemize}

\subsection{Device Numbers {\tt IDVOUT} and {\tt IDVERR}
\label{subsec:IDVOUT}}
The variables {\tt IDVOUT} and {\tt IDVERR} specify device numbers 
for ``running output" and ``error messages", respectively.
Normally these would both be set to the device number
for ``standard output" (e.g., writing to the screen if running interactively).
The variables {\tt IDVOUT} and {\tt IDVERR} are 
set by {\tt DATA} statements in the ``main" program
{\tt DDSCAT.f} and in the output routine {\tt WRIMSG} (file {\tt wrimsg.f}).  
Under Sun Fortran, {\tt DATA IDVOUT/0/} 
results in unbuffered output to
``standard output"; unbuffered output (if available) is nice so that if you
choose to direct your output to a file (e.g., using 
{\tt ddscat >\& ddscat.out \&})
the output file will contain up-to-date information.  Other operating systems
or compilers may not support this, and you may need to edit {\tt DDSCAT.f} to
change the two data statements to {\tt DATA IDVOUT/6/} and 
{\tt DATA IDVERR/6/}, and edit {\tt wrimsg.f} to change {\tt DATA IDVOUT/0/}
to {\tt DATA IDVOUT/6/}.

\subsection{Subroutine {\tt TIMEIT}\label{subsec:timeit}}
The only other operating system-dependent part of \ddscat\ is 
the single subroutine {\tt TIMEIT}.  
Several versions of {\tt TIMEIT} are provided:
\begin{itemize}
\item {\tt timeit\_sun.f} uses the SunOS system call {\tt etime}
\item {\tt timeit\_convex.f} uses the Convex OS system call {\tt etime}
\item {\tt timeit\_cray.f} uses the system call {\tt second}
\item {\tt timeit\_hp.f} uses the HP-AUX system calls {\tt sysconf} and 
{\tt times}
\item {\tt timeit\_ibm6000.f} uses the AIX system call {\tt mclock}
\item {\tt timeit\_osf.f} uses the DEC OSF system call {\tt etime}
\item {\tt timeit\_sgi.f} uses the IRIX system call {\tt etime}
\item {\tt timeit\_vms.f} uses the VMS system calls 
{\tt LIB\$INIT\_TIMER} and {\tt LIB\$SHOW\_TIMER}
\item {\tt timeit\_titan.f} uses the system call {\tt cputim}
\item {\tt timeit\_null.f} is a dummy routine which provides 
no timing information, but can be used under any operating system.
\end{itemize}
You {\it must} compile and link one of the 
{\tt timeit\_}{\it xxx}{\tt .f}
routines, possibly after modifying it to work on your system; 
{\tt timeit\_null.f}
is the easiest choice, but it will return no timing information.\footnote{
	Non-Unix sites: The VMS-compatible version of {\tt TIMEIT} is
	included in the file {\tt SRC9.FOR}.  For non-VMS sites, you will
	need to replace this version of {\tt TIMEIT} with one of the
	other versions, which are to be found in the file {\tt MISC.FOR}.
	If you are having trouble getting this to work, choose the
	``dummy'' version of {\tt TIMEIT} from {\tt MISC.FOR}: this will
	return no timing information, but is platform-independent.
	}

\subsection{Optimization}

The performance of \ddscat\ will benefit from optimization during compilation
and the user should enable the
appropriate compiler flags 
(e.g., \ {\tt g77 -c -O}, or \ {\tt pgf77 -c -fast}).

There is one exception: the {\tt LAPACK} routine {\tt SLAMC1}, which is called
to determine the number of bits of precision provided by the computer
architectures, should {\bf NOT} be optimized, because under some optimizers
the resulting code will end up in an endless loop.
Therefore the Makefile has a separate compilation flag 
{\tt fFLAGSslamc1} which should {\it not} specify optimization
(e.g., just \ {\tt g77 -c}, or \ {\tt pgf77 -c}).
\footnote{Note that the {\tt LAPACK} routines are only used for some target 
	geometries, and even when used the computation time is insignificant.}

\subsection{Leaving {\tt FFTW} Capability Disabled}

{\tt FFTW}\footnote{\tt http://www.fftw.org} 
-- ``Fastest Fourier Transform in the West'' -- is a publicly
available FFT implementation that performs very well
(see \S\ref{sec:choice_of_fft}).
\ddscat\ supports use of {\tt FFTW}, but it is recommended
that first-time users of {\tt DDSCAT} first get the code up and running
without {\tt FFTW} support, using the built-in GPFA FFT routine written
by Clive Temperton, which performs almost as well as FFTW.
This is also the option you will have to follow if {\tt FFTW} is not
installed on your system (though you can install it yourself, if necessary --
see \S\ref{subsec:support_for_FFTW} below).

The Makefile supplied with the \ddscat\ distribution is initially
set up to leave support for {\tt FFTW} disabled by
using a ``dummy'' subroutine {\tt dummycxfftw.f}. 

When you type {\tt make ddscat} you will create a version of {\tt ddscat}
which will not recognize option {\tt FFTW21} -- you {\it must} specify option
{\tt GFPAFT}.

\subsection{Enabling {\tt FFTW} Capability \label{subsec:support_for_FFTW}}

The {\tt FFTW} (``Fastest Fourier Transform in the West'') 
package of Matteo Frigo and Steven G. 
Johnson appears to be one of the
fastest public-domain FFT packages available (see \S\ref{sec:choice_of_fft}
for a performance comparison), 
and \ddscat\ allows this package to be used.  
In order to support the {\tt FFTW21}
option, however, it is necessary to first install the FFTW package on your
system (FFTW 2.1.x -- we have not yet implemented support for FFTW 3.0,
which has a different interface).

\begin{figure}
\centerline{\includegraphics[width=9.0cm]{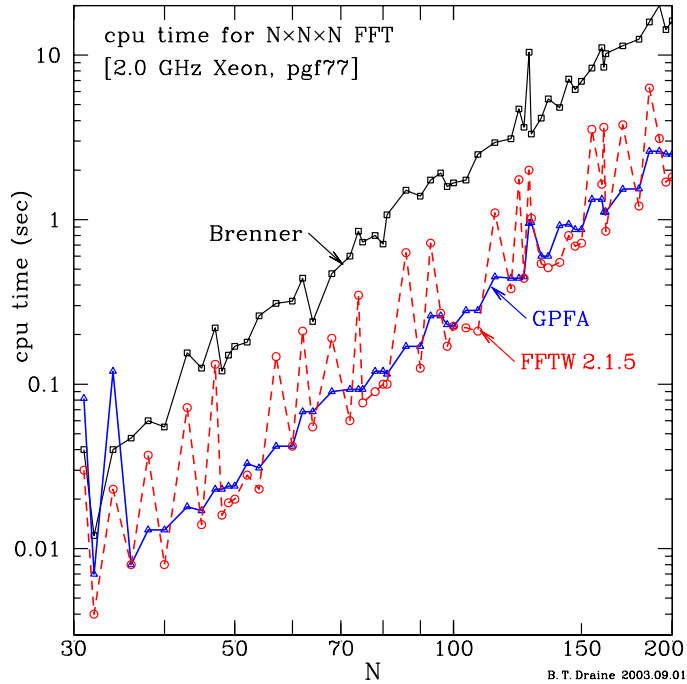}}
\caption{\label{fig:fft_timings}
	Comparison of cpu time required by 3 different FFT
	implementations.
	It is seen that the GPFA and FFTW implementations have comparable
	speeds, much faster than Brenner's FFT implementation.
	}
\end{figure}

{\tt FFTW} offers two advantages:
\begin{itemize}
\item It appears, for some cases, 
	to be somewhat faster than the GPFA FFT algorithm which
	is already included in the {\tt DDSCAT} package
	(see \S\ref{sec:choice_of_fft} and Fig.\ \ref{fig:fft_timings}).
\item It does not restrict the dimensions of the ``computational volume''
	to be factorizable as $2^i3^j5^k$, and therefore for some
	targets will use a smaller computational volume (and therefore
	require less memory).
\end{itemize}

If the {\tt FFTW} package is not yet installed on your system, then
\begin{enumerate}
\item Download the {\tt FFTW 2.1.x} package from {\tt http://www.fftw.org} 
	(as of this
	writing the latest version 2.1.x is {\tt fftw-2.1.5.tar.gz}) into
	some convenient directory.
\item {\tt tar xvfz fftw-2.1.5.tar.gz}
\item {\tt cd fftw-2.1.5}
\item {\tt ./configure --enable-float --enable-type-prefix --prefix=PATH}\\
	where {\tt PATH} is the path to the directory in which you would
	like the fftw library installed (you must have write permission,
	of course).
	If you choose to omit \\
	\hspace*{4em}{\tt --prefix=PATH} \\
	then fftw will be installed
	in the default directory {\tt /usr/local/}, but unless you are
	root you may not have write permission to this directory.
\item {\tt make}
\item {\tt make install}
\end{enumerate}
This should install the file {\tt libsfftw.a} in the directory
{\tt /usr/local/lib} (or {\tt PATH/lib} if you used option
{\tt --prefix=PATH} when running {\tt configure}).

Once FFTW is installed, you will need to edit the Makefile:
\begin{itemize}
\item define {\tt cxfftw = cxfftw}
\item define {\tt LIBFFTW} to give the correct path to {\tt libsfftw.a}
\end{itemize}

Once the above is done, typing {\tt make ddscat} should create an
executable {\tt ddscat} which will recognize option {\tt FFTW21} in
the file {\tt ddscat.par}.  Note that you will be linking fortran to C,
and different compilers may behave idiosyncratically.
The Makefile has examples which work for {\tt g77} and {\tt pgf77}.
If you have trouble, consult with someone familiar with linking fortran
and C modules on your system.

\subsection{Leaving the netCDF Capability Disabled}

{\tt netCDF}\footnote{http://www.unidata.ucasr.edu/packages/netcdf/} 
is a standard for data transport used at many sites 
(see \S\ref{subsec:netCDF} for more information on {\tt netCDF}).
The {\tt Makefile} supplied with the distribution of \ddscat\ is
set up to link to a ``dummy'' subroutine {\tt dummywritenet.f} instead
of subroutine {\tt writenet.f}, in order to minimize problems during
initial compilation and linking.
The ``dummy'' routine has no functionality, other than bailing out
with a fatal error message if the user makes the mistake of trying
to specify one of the netCDF options ({\tt ALLCDF} or {\tt ORICDF}).
First-time users of \ddscat\ should {\it not} try to use the
netCDF option -- simply skip this section, and specify option
{\tt NOTCDF} in {\tt ddscat.par}.
After successfully using \ddscat\, you can return to 
\S\ref{subsec:enabling_netCDF}
to try to enable the netCDF capability.

\subsection{Enabling the netCDF Capability\label{subsec:enabling_netCDF}}

Subroutine {\tt WRITENET} (file {\tt writenet.f}) provides the capability
to output 
binary data in the netCDF standard format (see \S\ref{subsec:netCDF}).
In order to use this routine (instead of {\tt dummywritenet.f}), 
it is necessary to take the following steps:\footnote{
	Non-UNIX sites:
	You need to replace the ``dummy'' version of {\tt SUBROUTINE WRITENET}
	in {\tt SRC1.FOR} with the version provided in {\tt MISC.FOR}.
	You will also need to consult your systems administrator to verify
	that the netCDF library has already been installed on your system,
	and to find out how to link to the library routines.
	}
\begin{enumerate}
\item Have netCDF already installed on your system (check with your
	system administrator).
\item Find out where {\tt netcdf.inc} is located and
	edit the {\tt include} statement in {\tt writenet.f} to specify the
	correct path to {\tt netcdf.inc}.
\item Find out where the {\tt libnetcdf.a} library is located, and edit
	the Makefile so that the variable {\tt LIBNETCDF} specifies the correct
	path to this library.
\item Edit the Makefile to ``comment out'' (with a {\tt\#} symbol in column 1) 
the line {\tt writenet = dummywritenet} and ``uncomment'' (remove the 
{\tt\#} symbol) the line
{\tt writenet = writenet} so that {\tt writenet.f} will be compiled
instead of the dummy routine {\tt dummywritenet.f}.
\end{enumerate}

\subsection{Compiling and Linking...}
After suitably editing the Makefile 
(while still positioned in {\tt DDA/src}) 
simply type\footnote{
	Non-UNIX sites: see \S\ref{subsec:nonunix}
	}\hfill\break
\indent\indent{\tt make ddscat}\hfill\break
which should create an executable file {\tt DDA/src/ddscat} .
If the default (as-provided) Makefile is used,
the {\tt g77} fortran compiler will be used to create an
executable which:
\begin{itemize}
\item will {\it not} have {\tt MPI} support;
\item will {\it not} have {\tt FFTW} support;
\item will {\it not} have {\tt netCDF} capability;
\item will contain timing instructions compatible with Linux, 
Solaris 1.x or 2.x, as well as several other versions of Unix.
\end{itemize}

To add {\tt FFTW} capability, see \S\ref{subsec:support_for_FFTW}.
To add {\tt netCDF} capability, see \S\ref{subsec:enabling_netCDF}.
To add {\tt MPI} support, see \S\ref{sec:MPI}.
To replace the Linux-compatible and Sun-compatible 
timing routine with another, see
\S\ref{subsec:timeit}.

\subsection{Installation on Non-Unix Systems\label{subsec:nonunix}}

{{\bf DDSCAT.5a}} is written in standard Fortran-77 plus the
{\tt DO ... ENDDO} extension which appears to be supported by all current
Fortran-77 compatible compilers.  
It is possible to run {{\bf DDSCAT}} on non-Unix systems.
If the Unix ``make'' utility is not available, here in brief is what needs to 
be accomplished:

All of the necessary Fortran code to compile and link {{\bf DDSCAT.5a}}
is included in the following files:
{\tt SRC0.FOR}, {\tt SRC1.FOR}, {\tt SRC2.FOR},
{\tt SRC3.FOR},
{\tt SRC4.FOR}, {\tt SRC5.FOR}, {\tt SRC6.FOR}, {\tt SRC7.FOR}, {\tt SRC8.FOR},
{\tt SRC9.FOR}, {\tt CGCOMMON.FOR},
{\tt GPFA.FOR}, {\tt LAPACK.FOR}, and {\tt PIM.FOR}.
There is an additional file {\tt MISC.FOR}, but this is not 
needed for ``basic'' use of the code (see below).

The main program {\tt DDSCAT} is found in {\tt SRC0.FOR}.  It calls a number
of subroutines, which are included in the other {\tt *.FOR} files.

Four of the subroutines exist in more than one version.
The ``default'' version of each is located in the file {\tt SRC1.FOR}:
\begin{itemize}
\item Select the version of {\tt SUBROUTINE CXFFTW} from {\tt SRC1.FOR}
	instead of the version in {\tt MISC.FOR}.  The resulting code
	will not support {\bf FFTW} capability.  If you later wish
	to add FFTW capability, you will first need to install the FFTW 
	library on your system (see \S\ref{subsec:support_for_FFTW}).
	After you have done so, you can use the
	version of {\tt SUBROUTINE CXFFTW} provided in {\tt MISC.FOR}.
\item Select the version of {\tt SUBROUTINE WRITENET} from {\tt SRC1.FOR}
	instead of the version in {\tt MISC.FOR}.  The resulting code
	will not support {\bf netCDF} capability.  (If {\bf netCDF}
	capability is required, you will need to install {\bf netCDF} libraries
	on your system --
	see \S\ref{subsec:enabling_netCDF}).
\item Select the version of {\tt SUBROUTINE C3DFFT} from {\tt SRC1.FOR}
	(do not use the version in {\tt MISC.FOR})
\item There is one version of {\tt SUBROUTINE TIMEIT} 
	included in {\tt SRC1.FOR}, and a number of additional versions 
	in {\tt MISC.FOR}.
Use the version from {\tt SRC1.FOR}, which begins

\begin{verbatim}
      SUBROUTINE TIMEIT(CMSGTM,DTIME)
C
C     timeit_null
C
C This version of timeit is a dummy which does not provide any
C timing information.
\end{verbatim}
	This version does not use any system calls, and therefore the
	code should compile and link without problems; however, when you
	run the code, it will not report any timing information reporting
	how much time was spent on different parts of the calculation.

	If you wish to obtain
	timing information, you will need to find out what system calls
	are supported by your operating system.  You can look at the
	other versions of {\tt SUBROUTINE TIMEIT} in {\tt MISC.FOR} 
	to see how this has
	been done under VMS and various version of Unix.
\end{itemize}
Once you have selected the appropriate versions of {\tt CXFFTW},
	{\tt WRITENET},
	{\tt C3DFFT}, and {\tt TIMEIT}, you
	can simply compile and link just as you would with any other Fortran
code with a number of modules.
You should end up with an executable with a (system-dependent) name
like {\tt DDSCAT.EXE}.

In addition to program {{\tt DDSCAT}}, we provide two other Fortran 77 
programs which may be useful.  Program {\tt CALLTARGET} can be used
to call the target generation routines which may be helpful for testing
purposes.
Program {\tt TSTFFT} is useful for comparing speeds of different FFT options.

\section{Moving the Executable}

Now reposition yourself into the directory {\tt DDA}\
(e.g., type {\tt cd ..}), 
and copy the executable
from {\tt src/ddscat} to the {\tt DDA}\ directory by typing

\indent\indent {\tt cp src/ddscat ddscat}

\noindent This should copy the file {\tt DDA/src/ddscat} to 
{\tt DDA/ddscat} (you could, of course, simply create a symbolic
link instead).  Similarly,
copy the sample parameter file {\tt ddscat.par} and the file 
{\tt diel.tab} to the
{\tt DDA} directory by typing

\indent\indent{\tt cp doc/ddscat.par ddscat.par}\hfill\break
\indent\indent{\tt cp doc/diel.tab diel.tab}

\section{The Parameter File {\tt ddscat.par}\label{sec:parameter_file}}

The directory {\tt DDA}\ should now contain a sample file 
{\tt ddscat.par} which
provides parameters to the program {\tt ddscat}.  
As provided (see Appendix\ref{app:ddscat.par}),
the file {\tt ddscat.par} 
is set up to calculate scattering by a 8$\times$6$\times$4 
rectangular array of 192
dipoles, with an effective radius $a_{\rm eff}=1{\mu{\rm m}}$, at a wavelength of 
$6.2832{\mu{\rm m}}$ (for a ``size parameter'' $2\pi a_{\rm eff}/\lambda=1$).

The dielectric function of the target material is provided in the file
{\tt diel.tab}, which is a sample file in which the refractive index is set to
$m=1.33+0.01i$ at all wavelengths; the name of this file is provided to 
{\tt ddscat} by
the parameter file {\tt ddscat.par}.

The sample parameter file as supplied calls 
for the GPFA FFT routine ({\tt GPFAFT}) 
of Temperton (1992) to be employed and the {\tt PBCGST} iterative method
to be used for solving the system of linear equations.
(See section \S\ref{sec:choice_of_fft} and \S\ref{sec:choice_of_algorithm} 
for discussion of choice of FFT algorithm and 
choice of equation-solving algorithm.)

The sample parameter file specifies (via option {\tt LATTDR}) that the 
``Lattice Dispersion Relation'' (Draine \& Goodman 1993)
be employed to determine the
dipole polarizabilities.
See \S\ref{sec:polarizabilities} for discussion of other options.

The sample parameter files specifies options {\tt NOTBIN} and {\tt NOTCDF}
so that no binary data files and no NetCDF files will be 
created by {\tt ddscat}.

The sample {\tt ddscat.par} file specifies that the calculations be done for a
single wavelength ($6.2832{\mu{\rm m}}$) and a single effective radius 
($1.00{\mu{\rm m}}$).
Note that in \ddscat\ the ``effective radius'' 
$a_{\rm eff}$ is the radius of a sphere of
equal volume -- i.e., a sphere of volume $Nd^3$ , where $d$ 
is the lattice spacing
and $N$ is the number of occupied (i.e., non-vacuum) 
lattice sites in the target.
Thus the effective radius $a_{\rm eff} = (3N/4\pi)^{1/3}d$ .

The incident radiation is always assumed to propagate along the $x$ axis in
the ``Lab Frame''.  
The sample {\tt ddscat.par} file specifies incident polarization
state ${\hat{\bf e}}_{01}$ to be along the $y$ axis 
(and consequently polarization state ${\hat{\bf e}}_{02}$
will automatically be taken to be along the $z$ axis).  
{\tt IORTH=2} in {\tt ddscat.par}
calls for calculations to be carried out for both incident polarization
states (${\hat{\bf e}}_{01}$ and ${\hat{\bf e}}_{02}$
-- see \S\ref{sec:incident_polarization}).

The target is assumed to have two vectors ${\hat{\bf a}}_1$ and
${\hat{\bf a}}_2$ embedded in it; ${\hat{\bf a}}_2$ is perpendicular
to ${\hat{\bf a}}_1$.  In the case of the 8$\times$6$\times$4
rectangular array of the sample calculation, the vector ${\hat{\bf
a}}_1$ is along the ``long'' axis of the target, and the vector
${\hat{\bf a}}_2$ is along the ``intermediate'' axis.  The target
orientation in the Lab Frame is set by three angles: $\beta$,
$\Theta$, and $\Phi$, defined and discussed below in
\S\ref{sec:target_orientation}.  Briefly, the polar angles $\Theta$
and $\Phi$ specify the direction of ${\hat{\bf a}}_1$ in the Lab
Frame.  The target is assumed to be rotated around ${\hat{\bf a}}_1$
by an angle $\beta$.  The sample {\tt ddscat.par} file specifies
$\beta=0$ and $\Phi=0$ (see lines in {\tt ddscat.par} specifying
variables {\tt BETA} and {\tt PHI}), and calls for three values of the
angle $\Theta$ (see line in {\tt ddscat.par} specifying variable {\tt
THETA}).  \ddscat\ chooses $\Theta$ values uniformly spaced
in $\cos\Theta$; thus, asking for three values of $\Theta$ between 0
and $90^\circ$ yields $\Theta=0$, $60^\circ$, and $90^\circ$.

Appendix \ref{app:ddscat.par} provides a detailed description of the 
file {\tt ddscat.par}.\footnote{
	Note that the structure of {\tt ddscat.par} depends on the version 
	of {{\bf DDSCAT}}\ being used.
	Make sure you update old parameter files before using them with
	\ddscat\ !
	}

\section{Running \ddscat\ Using the Sample {\tt ddscat.par} File}

To execute the program on a UNIX system (running either {\tt sh} 
or {\tt csh}), simply type

\indent\indent {\tt ddscat >\& ddscat.out \&}

\noindent which will redirect the ``standard output'' to the file {\tt
ddscat.out}, and run the calculation in the background.  The sample
calculation (8x6x4=192 dipole target, 3 orientations, two incident
polarizations, with scattering calculated for 14 distinct scattering
directions), requires 1.0 cpu seconds on a Sun Ultra-170 workstation.

\section{Output Files}

\subsection{ASCII files\label{subsec:ascii}}

If you run DDSCAT using the command\hfill\break \indent\indent{\tt
ddscat >\& ddscat.out \&} \hfill\break you will have various types of
ASCII files when the computation is complete:
\begin{itemize}
\item a file {\tt ddscat.out};
\item a file {\tt mtable};
\item a file {\tt qtable};
\item a file {\tt qtable2};
\item files {\tt w}{\it xx}{\tt r}{\it yy}{\tt ori.avg} 
	(one, {\tt w00r00ori.avg}, for the sample calculation);
\item if {\tt ddscat.par} specified {\tt IWRKSC}=1, there will also be
	files {\tt w}{\it xx}{\tt r}{\it yy}{\tt k}{\it zzz}{\tt .sca} (3 for
	the sample calculation: {\tt w00r00k000.sca}, {\tt w00r00k001.sca},
	{\tt w00r00k002.sca}).
\end{itemize}
The file {\tt ddscat.out} will contain any error messages generated as
well as a running report on the progress of the calculation, including
creation of the target dipole array.  During the iterative
calculations, $Q_{\rm ext}$, $Q_{\rm abs}$, and $Q_{pha}$ are printed after
each iteration; you will be able to judge the degree to which
convergence has been achieved.  Unless {\tt TIMEIT} has been disabled,
there will also be timing information.

The file {\tt mtable} contains a summary of the dielectric constant
used in the calculations.

The file {\tt qtable} contains a summary of the
orientationally-averaged values of $Q_{\rm ext}$, $Q_{\rm abs}$, $Q_{\rm sca}$,
$g(1)=\langle\cos(\theta_s)\rangle$, and $Q_{bk}$.  Here $Q_{\rm ext}$,
$Q_{\rm abs}$, and $Q_{\rm sca}$ are the extinction, absorption, and
scattering cross sections divided by $\pi a_{\rm eff}^2$.  $Q_{bk}$ is
the differential cross section for backscattering (area per sr)
divided by $\pi a_{\rm eff}^2$.

The file {\tt qtable2} contains a summary of the
orientationally-averaged values of $Q_{pha}$, $Q_{pol}$, and
$Q_{cpol}$.  Here $Q_{pha}$ is the ``phase shift'' cross section
divided by $\pi a_{\rm eff}^2$ (see definition in Draine 1988).
$Q_{pol}$ is the ``polarization efficiency factor'', equal to the
difference between $Q_{\rm ext}$ for the two orthogonal polarization
states.  We define a ``circular polarization efficiency factor''
$Q_{cpol}\equiv Q_{pol}Q_{pha}$, since an optically-thin medium with a
small twist in the alignment direction will produce circular
polarization in initially unpolarized light in proportion to
$Q_{cpol}$.

For each wavelength and size, \ddscat\ produces a file with a
name of the form\break{\tt w{\it xx}r{\it yy}ori.avg}, where index
{\it xx} (=00, 01, 02....)  designates the wavelength and index {\it
yy} (=00, 01, 02...) designates the ``radius''; this file contains $Q$
values and scattering information averaged over however many target
orientations have been specified (see \S\ref{sec:target_orientation}.
The file {\tt w00r00ori.avg} produced by the sample calculation is
provided below in Appendix \ref{app:w00r00ori.avg}.

In addition, if {\tt ddscat.par} has specified {\tt IWRKSC}=1 (as for
the sample calculation), \ddscat\ will generate files with
names of the form {\tt w{\it xx}r{\it yy}k{\it zzz}.avg}, where {\it
xx} and {\it yy} are as before, and index {\it zzz} =(000,001,002...)
designates the target orientation; these files contain $Q$ values and
scattering information for {\it each} of the target orientations.  The
structure of each of these files is very similar to that of the {\tt
w{\it xx}r{\it yy}ori.avg} files.  Because these files may not be of
particular interest, and take up disk space, you may choose to set
{\tt IWRKSC}=0 in future work.  However, it is suggested that you run
the sample calculation with {\tt IWRKSC}=1.

The sample {\tt ddscat.par} file specifies {\tt IWRKSC}=1 and calls
for use of 1 wavelength, 1 target size, and averaging over 3 target
orientations.  Running \ddscat\ with the sample {\tt
ddscat.par} file will therefore generate files {\tt w00r00k000.sca},
{\tt w00r00k001.sca}, and {\tt w00r00k002.sca} .  To understand the
information contained in one of these files, please consult Appendix
\ref{app:w00r00k000.sca}, which contains an example of the file {\tt
w00r00k000.sca} produced in the sample calculation.

\subsection{Binary Option\label{subsec:binary}}

It is possible to output an ``unformatted'' or ``binary'' file ({\tt
dd.bin}) with fairly complete information, including header and data
sections.  This is accomplished by specifying either {\tt ALLBIN} or
{\tt ORIBIN} in {\tt ddscat.par} .

Subroutine {\tt writebin.f} provides an example of how this can be
done.  The ``header'' section contains dimensioning and other
variables which do not change with wavelength, particle geometry, and
target orientation.  The header section contains data defining the
particle shape, wavelengths, particle sizes, and target orientations.
If {\tt ALLBIN} has been specified, the ``data'' section contains, for
each orientation, Mueller matrix results for each scattering
direction.  The data output is limited to actual dimensions of arrays;
e.g. {\tt nscat,4,4} elements of Muller matrix are written rather than
{\tt mxscat,4,4}.  This is an important consideration when writing
postprocessing codes.

A skeletal example of a postprocessing code was written by us (PJF)
and is provided in subdirectory {\tt DDA/IDL}.  If you do plan to use
the Interactive Data Language (IDL) for postprocessing, you may
consider using the netCDF binary file option which offers substantial
advantages over the FORTRAN unformatted write.  More information about
IDL is available at {\verb|http://www.rsinc.com/idl|}.  Unfortunately
IDL requires a license and hence is not distributed with {{\bf
DDSCAT}}.

\subsection{Machine-Independent Binary File Option: netCDF
	\label{subsec:netCDF}}

The ``unformatted'' binary file is compact, fairly complete, and (in
comparison to ASCII output files) is efficiently written from FORTRAN.
However, binary files are not compatible between different computer
architectures if written by ``unformatted'' {\tt WRITE} from FORTRAN.
Thus, you have to process the data on the same computer architecture
that was used for the {{\bf DDSCAT}}\ calculations.  We provide an
option of using netCDF with DDSCAT.  
The netCDF library\footnote{\label{fn:netCDF}
	{\tt http://www.unidata.ucar.edu/packages/netcdf/}}
defines a
machine-independent format for representing scientific data. Together,
the interface, library, and format support the creation, access, and
sharing of scientific data.
For more information, go to the {\tt netcdf} website$^{\ref{fn:netCDF}}$.
 
Several major graphics packages (for example IDL) have adopted netCDF
as a standard for data transport.  For this reason, and because of
strong and free support of netCDF over the network by UNIDATA, we have
implemented a netCDF interface in {{\bf DDSCAT}}.  This may become the
method of choice for binary file storage of output from {{\bf
DDSCAT}}.

After the initial ``learning curve'' netCDF offers many advantages: 
\begin{itemize}
\item It is easy to examine the contents of the file (using tools such as 
	{\tt ncdump}).
\item Binary files are portable - they  can be created on a supercomputer and 
	processed on a workstation.
\item Major graphics packages now offer netCDF interfaces.
\item Data input and output are an order of magnitude faster than for ASCII 
	files.
\item Output data files are compact.
\end{itemize}
The disadvantages include: 
\begin{itemize}
\item Need to have netCDF installed on your system.
\item Lack of support of complex numbers.
\item Nontrivial learning curve for using netCDF.
\end{itemize}

The public-domain netCDF software is available for many operating
systems from \break{\tt http://www.unidata.ucar.edu/packages/netcdf} .
The steps necessary for enabling the netCDF capability in {{\bf
DDSCAT.6.0}}\ are listed above in \S\ref{subsec:enabling_netCDF}.  Once
these have been successfully accomplished, and you are ready to run
{{\bf DDSCAT}}\ to produce netCDF output, you must choose either the
{\tt ALLCDF} or {\tt ORICDF} option in {\tt ddscat.par}; {\tt ALLCDF}
will result in a file which will include the Muller matrix for every
wavelength, particle size, and orientation, whereas {\tt ORICDF} will
result in a file limited to just the orientational averages for each
wavelength and target size.

\section{Dipole Polarizabilities\label{sec:polarizabilities}}

Option {\tt LATTDR} specifies that the 
``Lattice Dispersion Relation'' (Draine \& Goodman 1993)
be employed to determine the dipole
polarizabilities.; at this time this is the only supported option.

\section{Dielectric Functions\label{sec:dielectric_func}}

In order to assign the appropriate dipole polarizabilities, {{\bf
DDSCAT.6.0}}\ must be given the dielectric constant of the material (or
materials) of which the target of interest is composed.  Unless the
user wishes to use the dielectric function of either solid or liquid
H$_2$O (see below), his information is supplied to {{\bf DDSCAT}}\
through a table (or tables), read by subroutine {\tt DIELEC} in file
{\tt dielec.f}, and providing either the complex refractive index
$m=n+ik$ or complex dielectric function
$\epsilon=\epsilon_1+i\epsilon_2$ as a function of wavelength
$\lambda$.  Since $m=\epsilon^{1/2}$, or $\epsilon=m^2$, the user must
supply either $m$ or $\epsilon$, but not both.

The table formatting is intended to be quite flexible.  The first line
of the table consists of text, up to 80 characters of which will be
read and included in the output to identify the choice of dielectric
function.  (For the sample problem, it consists of simply the
statement {\tt m = 1.33 + 0.01i}.)  The second line consists of 5
integers; either the second and third {\it or} the fourth and fifth
should be zero.
\begin{itemize}
\item The first integer specifies which column the wavelength is stored in.
\item The second integer specifies which column Re$(m)$ is stored in.
\item The third integer specifies which column Im$(m)$ is stored in.
\item The fourth integer specifies which column Re$(\epsilon)$ is stored in.
\item The fifth integer specifies which column Im$(\epsilon)$ is stored in.
\end{itemize}
If the second and third integers are zeros, then {\tt DIELEC} will
read Re$(\epsilon)$ and Im$(\epsilon)$ from the file; if the fourth
and fifth integers are zeros, then Re$(m)$ and Im$(m)$ will be read
from the file.

The third line of the file is used for column headers, and the data
begins in line 4.  {\it There must be at least 3 lines of data:} even
if $\epsilon$ is required at only one wavelength, please supply two
additional ``dummy'' wavelength entries in the table so that the
interpolation apparatus will not be confused.

In the event that the user wishes to compute scattering by targets
with the refractive index of either solid or liquid H$_2$O, we have
included two ``built-in'' dielectric functions.  If {\tt H2OICE} is
specified on line 10 of {\tt ddscat.par}, {{\bf DDSCAT}}\ will use the
dielectric function of H$_2$O ice at T=250K as compiled by Warren
(1984).  If {\tt H2OLIQ} is specified on line 10 of {\tt ddscat.par},
{{\bf DDSCAT}}\ will use the dielectric function for liquid H$_2$O at
T=280K using subroutine {\tt REFWAT} written by Eric A. Smith.  For
more information see {\tt
http://atol.ucsd.edu/$\sim$pflatau/refrtab/index.htm} .

\section{Choice of FFT Algorithm\label{sec:choice_of_fft}}

One major change in going from {{\bf DDSCAT}}{\bf .4b} to {\bf 4c}
was modification of the code to permit use of the
GPFA FFT algorithm developed by Dr. Clive Temperton (1992).  
In going from {{\bf DDSCAT.5a10}}
to {{6.0}} we introduce a new FFT option: the {\tt FFTW} package
of Frigo and Johnson.
DDSCAT continues
to offer Temperton's GPFA code
as an alternative FFT implementations.  

To compare the performance of the GPFA algorithm
and the FFTW code,
we provide a program {\tt TSTFFT} to compare the performance
of different 3-D FFT implementations.  
The {\tt TSTFFT} code requires
the {\tt FFTW} library (see \S\ref{subsec:support_for_FFTW}).
{\tt TSTFFT} also provides timings for an FFT implementation due to
Brenner (included in previous releases of {\bf DDSCAT}).

To compile, link, and run this
program on a Unix system,\footnote{ Non-Unix systems: the {\tt TSTFFT}
Fortran source code is in the file {\tt MISC.FOR}.}  position yourself
in the {\tt DDA/src} directory and type

\indent\indent {\tt make tstfft}\hfill\break
\indent\indent {\tt tstfft}

\noindent Output results will be written into a file {\tt res.dat}.
Table \ref{tab:fftres} 
is the {\tt res.dat} file created when run on a 2.0 GHz 
Intel Xeon
system, using the pgf77 compiler:
\begin{table}
\caption{FFT timing output from program {\tt tstfft}\label{tab:fftres}}
{\scriptsize
\begin{verbatim}
 CPU time (sec) for different 3-D FFT methods
 Machine=2.0 GHz Xeon, 256kB L2 cache, pgf77 compiler 
 parameter LVR = 64 
 LVR="length of vector registers" in gpfa2f,gpfa3f,gpfa5f
============================================================
                  Brenner    Temperton Frigo & Johnson
   NX  NY  NZ     (FOURX)     (GPFA)       (FFTW)
   31  31  31       0.040       0.082       0.030
   32  32  32       0.012       0.007       0.004
   34  34  34       0.040       0.120       0.023
   36  36  36       0.047       0.008       0.008
   38  38  38       0.060       0.013       0.037
   40  40  40       0.055       0.013       0.008
   43  43  43       0.155       0.018       0.072
   45  45  45       0.125       0.017       0.014
   47  47  47       0.220       0.023       0.132
   48  48  48       0.120       0.023       0.016
   49  49  49       0.150       0.024       0.019
   50  50  50       0.170       0.024       0.020
   52  52  52       0.180       0.033       0.028
   54  54  54       0.260       0.031       0.023
   57  57  57       0.310       0.042       0.147
   60  60  60       0.320       0.042       0.042
   62  62  62       0.440       0.068       0.210
   64  64  64       0.240       0.068       0.055
   68  68  68       0.470       0.090       0.190
   72  72  72       0.600       0.093       0.060
   74  74  74       0.850       0.093       0.347
   75  75  75       0.730       0.093       0.077
   78  78  78       0.800       0.120       0.090
   80  80  80       0.710       0.120       0.100
   81  81  81       1.070       0.115       0.100
   86  86  86       1.510       0.170       0.630
   90  90  90       1.390       0.170       0.125
   93  93  93       1.740       0.260       0.720
   96  96  96       1.920       0.260       0.270
   98  98  98       1.590       0.230       0.170
  100 100 100       1.670       0.225       0.225
  104 104 104       1.740       0.280       0.220
  108 108 108       2.490       0.280       0.210
  114 114 114       2.940       0.450       1.100
  120 120 120       3.100       0.440       0.380
  123 123 123       4.710       0.440       1.750
  125 125 125       3.640       0.450       0.440
  127 127 127      10.400       0.950       2.000
  128 128 128       3.310       0.960       1.020
  132 132 132       4.140       0.600       0.540
  135 135 135       5.420       0.600       0.510
  140 140 140       4.800       0.920       0.550
  144 144 144       7.150       0.940       0.800
  147 147 147       6.170       0.870       0.690
  150 150 150       6.930       0.870       0.720
  155 155 155       8.350       1.330       3.540
  160 160 160      11.110       1.330       1.640
  161 161 161       8.450       1.110       3.640
  162 162 162      10.190       1.110       0.850
  171 171 171      11.380       1.530       3.770
  180 180 180      12.480       1.540       1.210
  186 186 186      15.920       2.600       6.310
  192 192 192      20.050       2.610       3.110
  196 196 196      14.270       2.500       1.690
  200 200 200      16.180       2.500       1.820
\end{verbatim}
}
\end{table}

It is clear that both the GPFA code and the FFTW code are 
{\it much} faster than the
Brenner FFT (by a factor of 7 for the 100$\times$100$\times$100 case).
The FFTW code and GPFA code are quite comparable in
performance -- for some cases the GPFA code is faster, for other cases
the FFTW code is faster.
For target dimensions which are factorizable as $2^i3^j5^k$ (for integer
$i$, $j$, $k$), the GPFA
and FFTW 
codes have the same memory requirements.
For targets with extents $N_x$, $N_y$, $N_z$ 
which are not factorizable as $2^i3^j5^k$, the GPFA code
needs to ``extend'' the computational volume to have values of $N_x$,
$N_y$, and $N_z$ which are factorizable by 2, 3, and 5.
For these cases, GPFA requires somewhat more memory than
FFTW.
However, the fractional difference in required memory 
is not large, since integers factorizable
as $2^i3^j5^k$ occur fairly 
frequently.\footnote{2, 3, 4, 5, 6, 8, 9, 10, 12,
	15, 16, 18, 20, 24, 25, 27, 30, 32, 36, 40,
	45, 48, 50, 54, 60, 64, 72, 75, 80, 81, 90, 
	96, 100, 108, 120, 125, 128, 135,
	144, 150, 160, 162, 180, 192, 200 are the integers $\leq 200$
	which are of the form $2^i3^j5^k$.}

Based on Figure 5, we see that while for some cases {\tt FFTW 2.1.5} is
faster than the GPFA algorithm, the difference appears to be fairly marginal.

The GPFA code contains a parameter {\tt LVR} which is set in {\tt
data} statements in the routines {\tt gpfa2f}, {\tt gpfa3f}, and {\tt
gpfa5f}.  {\tt LVR} is supposed to be optimized to correspond to the
``length of a vector register'' on vector machines.  As delivered,
this parameter is set to 64, which is supposed to be appropriate for
Crays other than the C90 (for the C90, 128 is supposed to be
preferable).\footnote{ 
	In place of ``preferable'' users are encouraged
	to read ``necessary'' -- there is some basis for fearing that results
	computed on a C90 with {\tt LVR} other than 128 run the risk of being
	incorrect!  Please use {\tt LVR=128} if running on a C90!
	}
The value of {\tt LVR} is not critical for scalar machines, as long as
it is fairly large.  We found little difference between {\tt LVR}=64
and 128 on a Sparc 10/51, on an Ultrasparc 170, and on an Intel Xeon cpu.
You may wish to
experiment with different {\tt LVR} values on your computer
architecture.  To change {\tt LVR}, you need to edit {\tt gpfa.f} and
change the three {\tt data} statements where {\tt LVR} is set.

The choice of FFT implementation is obtained by specifying one of:
\begin{itemize}
\item{\tt FFTW21} to use the FFTW 2.1.x algorithm of Frigo and Johnson --
	{\bf this is recommended, but requires that the FFTW 2.1.x library be
	installed on your system}.
\item{\tt GPFAFT} to use the GPFA algorithm (Temperton 1992) -- {\bf this is
	almost as fast as the FFTW package, and does not require that the
	FFTW package be installed on your system};
\item{\tt CONVEX} to use the native FFT routine on a Convex system
	(this option is presumably obsolete, but is retained as an example
	of how to use a system-specific native routine).
\end{itemize}

\section{Choice of Iterative Algorithm\label{sec:choice_of_algorithm}}

As discussed elsewhere (e.g., Draine 1988), the problem of
electromagnetic scattering of an incident wave ${\bf E}_{\rm inc}$ by an
array of $N$ point dipoles can be cast in the form
\begin{equation}
{\bf A} {\bf P} = {\bf E}
\label{eq:AP=E}
\end{equation}
where ${\bf E}$ is a $3N$-dimensional (complex) vector of the incident
electric field ${\bf E}_{\rm inc}$ at the $N$ lattice sites, ${\bf P}$ is
a $3N$-dimensional (complex) vector of the (unknown) dipole
polarizations, and ${\bf A}$ is a $3N\times3N$ complex matrix.

Because $3N$ is a large number, direct methods for solving this system
of equations for the unknown vector ${\bf P}$ are impractical, but
iterative methods are useful: we begin with a guess (typically, ${\bf
P}=0$) for the unknown polarization vector, and then iteratively
improve the estimate for ${\bf P}$ until equation (\ref{eq:AP=E}) is
solved to some error criterion.  The error tolerance may be specified
as
\begin{equation}
{|{\bf A}^\dagger {\bf A} {\bf P} - {\bf A}^\dagger {\bf E}| \over
| {\bf A}^\dagger {\bf E} |}
 < h 
\label{eq:err_tol}~~~,
\end{equation}
where ${\bf A}^\dagger$ is the Hermitian conjugate of ${\bf A}$
[$(A^\dagger)_{ij} \equiv (A_{ji})^*$], and $h$ is the error
tolerance.  We typically use $h=10^{-5}$ in order to satisfy
eq.(\ref{eq:AP=E}) to high accuracy.  The error tolerance $h$ can be
specified by the user (see Appendix \ref{app:ddscat.par}).

A major change in going from {{\bf DDSCAT}}{\bf .4b} to {\bf 5a} (and
subsequent versions) was the implementation of several different
algorithms for iterative solution of the system of complex linear
equations.  {{\bf DDSCAT.5a}}\ and \ddscat\ have been
structured to permit solution algorithms to be treated in a fairly
``modular'' fashion, facilitating the testing of different algorithms.
Many algorithms were compared by Flatau (1997)\footnote{ A postscript
copy of this report -- file {\tt cg.ps} -- is distributed with the
\ddscat\ documentation.}; two of them performed well and are
made available to the user in \ddscat.  The choice of
algorithm is made by specifying one of the options:
\begin{itemize}
\item {\tt PBCGST} -- Preconditioned BiConjugate Gradient with 
	STabilitization method from the Parallel Iterative Methods 
	(PIM) package created by R. Dias da Cunha and T. Hopkins.
\item {\tt PETRKP} -- the complex conjugate gradient algorithm of 
	Petravic \& Kuo-Petravic (1979), as coded in the Complex Conjugate 
	Gradient package (CCGPACK) created by P.J. Flatau.
	This is the algorithm discussed by Draine (1988) and used in 
	previous versions of {{\bf DDSCAT}}.
\end{itemize}
Both methods work well.  Our experience suggests that {\tt PBCGST} is
often faster than {\tt PETRKP}, by perhaps a factor of two.  We
therefore recommend it, but include the {\tt PETRKP} method as an
alternative.

The Parallel Iterative Methods (PIM) by Rudnei Dias da Cunha ({\tt
rdd@ukc.ac.uk}) and Tim Hopkins ({\tt trh@ukc.ac.uk}) is a collection
of Fortran 77 routines designed to solve systems of linear equations
on parallel and scalar computers using a variety of iterative methods
(available at \hfill\break {\tt http://www.mat.ufrgs.br/pim-e.html}).
PIM offers a number of iterative methods, including
\begin{itemize}
\item Conjugate-Gradients, CG (Hestenes 1952), 
\item Bi-Conjugate-Gradients, BICG (Fletcher 1976), 
\item Conjugate-Gradients squared, CGS (Sonneveld 1989), 
\item the stabilised version of Bi-Conjugate-Gradients, BICGSTAB (van
	der Vorst 1991),
\item the restarted version of BICGSTAB, RBICGSTAB (Sleijpen \& Fokkema 1992) 
\item the restarted generalized minimal residual, RGMRES (Saad 1986), 
\item the restarted generalized conjugate residual, RGCR (Eisenstat 1983), 
\item the normal equation solvers, CGNR (Hestenes 1952 and CGNE (Craig 1955), 
\item the quasi-minimal residual, QMR (highly parallel version due to 
	Bucker \& Sauren 1996), 
\item transpose-free quasi-minimal residual, TFQMR (Freund 1992), 
\item the Chebyshev acceleration, CHEBYSHEV (Young 1981). 
\end{itemize}
The source code for these methods is distributed with {\tt DDSCAT} but
only {\tt PBCGST} and {\tt PETRKP} can be called directly via {\tt
ddscat.par}. It is possible (and was done) to add other options by
changing the code in {\tt getfml.f} .  A helpful introduction to
conjugate gradient methods is provided by the report ``Conjugate
Gradient Method Without Agonizing Pain" by Jonathan R. Shewchuk,
available as a postscript file: {\tt
ftp://REPORTS.ADM.CS.CMU.EDU/usr0/anon/1994/CMU-CS-94-125.ps}.

\section{Calculation of Radiative Force and Torque
	\label{sec:torque_calculation}}

In addition to solving the scattering problem for a dipole array,
\ddscat\ can compute the three-dimensional force ${\bf
F}_{\rm rad}$ and torque ${\bf \Gamma}_{\rm rad}$ exerted on this array by the
incident and scattered radiation fields.  This calculation is carried
out, after solving the scattering problem, provided {\tt DOTORQ} has
been specified in {\tt ddscat.par}.  For each incident polarization
mode, the results are given in terms of dimensionless efficiency
vectors ${\bf Q}_{pr}$ and ${\bf Q}_{\Gamma}$, defined by
\begin{equation}
{\bf Q}_{pr} \equiv {{\bf F}_{\rm rad} \over 
\pi a_{\rm eff}^2 u_{\rm rad}} ~~~,
\end{equation}
\begin{equation}
{\bf Q}_\Gamma \equiv {k{\bf \Gamma}_{\rm rad} \over
\pi a_{\rm a_eff}^2 u_{\rm rad}} ~~~,
\end{equation}
where ${\bf F}_{\rm rad}$ and ${\bf \Gamma}_{\rm rad}$ are the time-averaged
force and torque on the dipole array, $k=2\pi/\lambda$ is the
wavenumber {\it in vacuo}, and $u_{\rm rad} = E_0^2/8\pi$ is the
time-averaged energy density for an incident plane wave with amplitude
$E_0 \cos(\omega t + \phi)$.  The radiation pressure efficiency vector
can be written
\begin{equation}
{\bf Q}_{\rm pr} = Q_{\rm ext}{\hat{\bf k}} - Q_{\rm sca}{\bf g} ~~~,
\end{equation}
where ${\hat{\bf k}}$ is the direction of propagation of the incident
radiation, and the vector {\bf g} is the mean direction of propagation
of the scattered radiation:
\begin{equation}
{\bf g} = {1\over C_{\rm sca}}\int d\Omega
{dC_{\rm sca}({\hat{\bf n}},{\hat{\bf k}})\over d\Omega} {\hat{\bf n}} ~~~,
\end{equation}
where $d\Omega$ is the element of solid angle in scattering direction
${\hat{\bf n}}$, and $dC_{\rm sca}/d\Omega$ is the differential scattering
cross section.

Equations for the evaluation of the radiative force and torque are
derived by Draine \& Weingartner (1996).  It is important to note that
evaluation of ${\bf Q}_{pr}$ and ${\bf Q}_\Gamma$ involves averaging
over scattering directions to evaluate the linear and angular momentum
transport by the scattered wave.  This evaluation requires appropriate
choices of the parameters {\tt ICTHM} and {\tt IPHM} -- see
\S\ref{sec:averaging_scattering}.

\section{Memory Requirements \label{sec:memory_requirements}}

Since Fortran-77 does not allow for dynamic memory allocation, the
executable has compiled into it the dimensions for a number of arrays;
these array dimensions limit the size of the dipole array which the
code can handle.  The source code supplied to you (which can be used
to run the sample calculation with 192 dipoles) is restricted to
problems with targets with a maximum extent of 16 lattice spacings
along the $x$-, $y$-, and $z$-directions ({\tt MXNX=16,MXNY=16,MXNZ=16};
i.e, the target must fit within an 16$\times$16$\times$16=4096 cube) and
involve at most 9 different dielectric functions ({\tt MXCOMP=9}).
With this dimensioning, the executable requires about 1.3 MB of memory
to run on an Ultrasparc system; memory requirements on other hardware
and with other compilers should be similar.  To run larger problems,
you will need to edit the code to change {\tt PARAMETER} values and
recompile.

All of the dimensioning takes place in the main program {\tt DDSCAT}
-- this should be the only routine which it is necessary to recompile.
All of the dimensional parameters are set in {\tt PARAMETER}
statements appearing before the array declarations.  You need simply
edit the parameter statements.  Remember, of course, that the amount
of memory allocated by the code when it runs will depend upon these
dimensioning parameters, so do not set them to ridiculously large
values!  The parameters {\tt MXNX}, {\tt MXNY}, {\tt MXNZ} specify the
maximum extent of the target in the $x$-, $y$-, or $z$-directions.
The parameter {\tt MXCOMP} specifies the maximum number of different
dielectric functions which the code can handle at one time.  The
comment statements in the code supply all the information you should
need to change these parameters.

The memory requirement for \ddscat\ (with the netCDF and FFTW 
options disabled) is approximately 
\begin{equation}
(1690+0.764{\tt MXNX}\times{\tt
MXNY}\times{\tt MXNZ}) {\rm ~kbytes}
\end{equation}
Thus a
32$\times$32$\times$32 calculation requires 22.5 MBytes, while a
48$\times$48$\times$48 calculation requires 73.0 MBytes.  These values
are for the {\tt pgf77} compiler on an Intel Xeon system
running the Linux operating system.

\section{Target Orientation \label{sec:target_orientation}}

Recall that we define a ``Lab Frame'' (LF) in which the incident
radiation propagates in the $+x$ direction.  In {\tt ddscat.par} one
specifies the first polarization state ${\hat{\bf e}}_{01}$ (which
obviously must lie in the $y,z$ plane in the LF); {{\bf DDSCAT}}\
automatically constructs a second polarization state ${\hat{\bf
e}}_{02} = {\hat{\bf x}} \times {\hat{\bf e}}_{01}^*$ orthogonal to
${\hat{\bf e}}_{01}$ (here ${\hat{\bf x}}$ is the unit vector in the
$+x$ direction of the LF.  For purposes of discussion we will always
let unit vectors ${\hat{\bf x}}$, ${\hat{\bf y}}$, ${\hat{\bf
z}}={\hat{\bf x}}\times{\hat{\bf y}}$ be the three coordinate axes of
the LF.  Users will often find it convenient to let polarization
vectors ${\hat{\bf e}}_{01}={\hat{\bf y}}$, ${\hat{\bf
e}}_{02}={\hat{\bf z}}$ (although this is not mandatory -- see
\S\ref{sec:incident_polarization}).

\begin{figure}
\centerline{\includegraphics[width=16.cm]{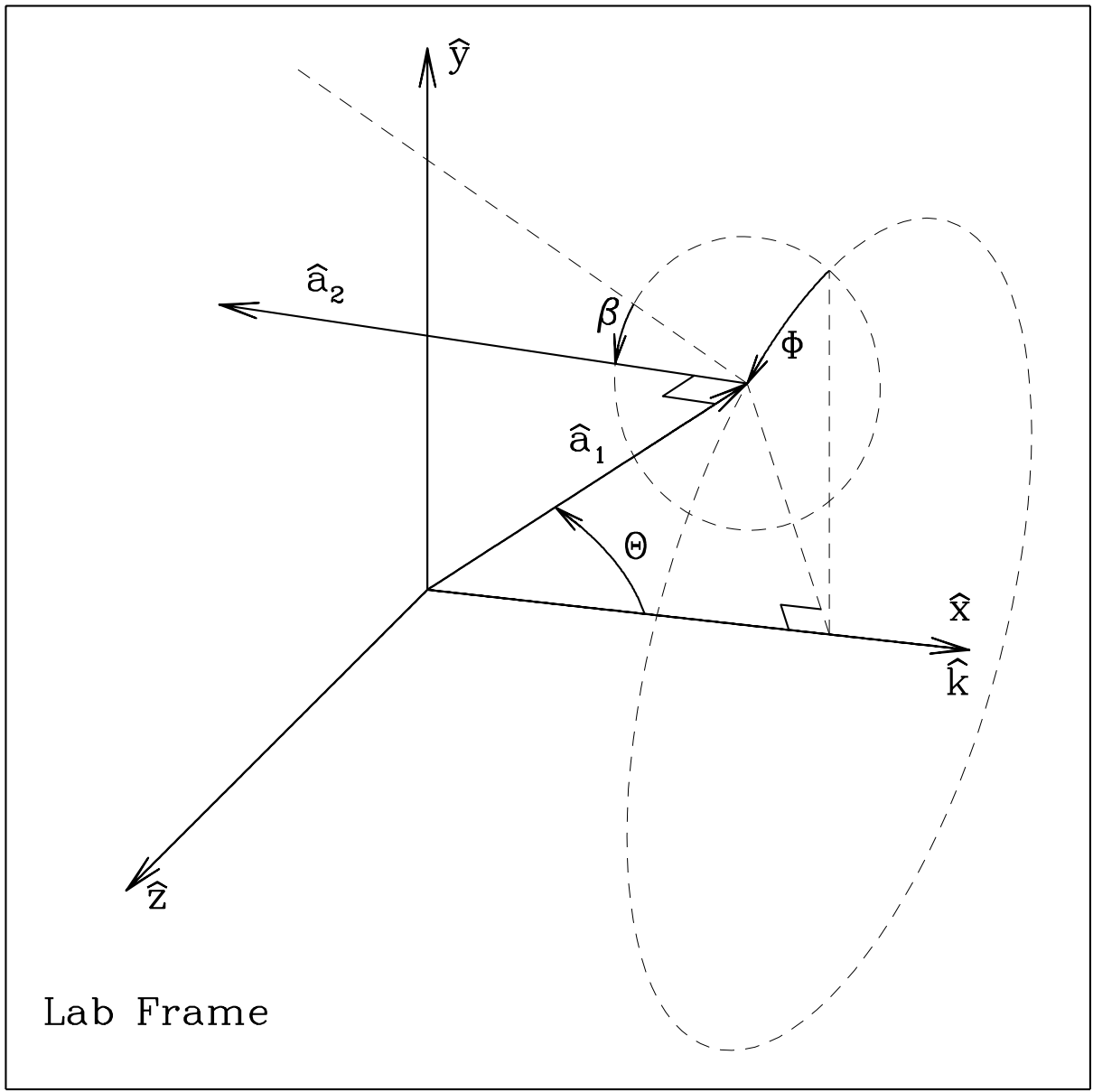}}
\caption{Target orientation in the Lab Frame.  ${\hat{\bf x}}$ is the
	direction of propagation of the incident radiation, and ${\hat{\bf
	y}}$ is the direction of the ``real'' component of the first incident
	polarization mode.  In this coordinate system, the orientation of
	target axis ${\hat{\bf a}}_1$ is specified by angles $\Theta$ and
	$\Phi$.  With target axis ${\hat{\bf a}}_1$ fixed, the orientation of
	target axis ${\hat{\bf a}}_2$ is then determined by angle $\beta$
	specifying rotation of the target around ${\hat{\bf a}}_1$.  When
	$\beta=0$, ${\hat{\bf a}}_2$ lies in the ${\hat{\bf a}}_1$,${\hat{\bf
	x}}$ plane.
	}
\label{fig:target_orientation}
\end{figure}

Recall that definition of a target involves specifying two unit
vectors, ${\hat{\bf a}}_1$ and ${\hat{\bf a}}_2$, which are imagined
to be ``frozen'' into the target.  We require ${\hat{\bf a}}_2$ to be
orthogonal to ${\hat{\bf a}}_1$.  Therefore we may define a ``Target
Frame" (TF) defined by the three unit vectors ${\hat{\bf a}}_1$,
${\hat{\bf a}}_2$, and ${\hat{\bf a}}_3 = {\hat{\bf a}}_1 \times
{\hat{\bf a}}_2$ .

For example, when {{\bf DDSCAT}}\ creates a 8$\times$6$\times$4
rectangular solid, it fixes ${\hat{\bf a}}_1$ to be along the longest
dimension of the solid, and ${\hat{\bf a}}_2$ to be along the
next-longest dimension.

Orientation of the target relative to the incident radiation can in
principle be determined two ways:
\begin{enumerate}
\item specifying the direction of ${\hat{\bf a}}_1$ and ${\hat{\bf
	a}}_2$ in the LF, or
\item specifying the directions of ${\hat{\bf x}}$ (incidence
	direction) and ${\hat{\bf y}}$ in the TF.
\end{enumerate}
\ddscat\ uses method 1.: the angles $\Theta$, $\Phi$, and
$\beta$ are specified in the file {\tt ddscat.par}.  The target is
oriented such that the polar angles $\Theta$ and $\Phi$ specify the
direction of ${\hat{\bf a}}_1$ relative to the incident direction
${\hat{\bf x}}$, where the ${\hat{\bf x}}$,${\hat{\bf y}}$ plane has
$\Phi=0$.  Once the direction of ${\hat{\bf a}}_1$ is specified, the
angle $\beta$ than specifies how the target is to rotated around the
axis ${\hat{\bf a}}_1$ to fully specify its orientation.  A more
extended and precise explanation follows:

\subsection{Orientation of the Target in the Lab Frame}

DDSCAT uses three angles, $\Theta$, $\Phi$, and $\beta$, to specify
the directions of unit vectors ${\hat{\bf a}}_1$ and ${\hat{\bf a}}_2$
in the LF (see Fig.\ \ref{fig:target_orientation}).

$\Theta$ is the angle between ${\hat{\bf a}}_1$ and ${\hat{\bf x}}$.

When $\Phi=0$, ${\hat{\bf a}}_1$ will lie in the ${\hat{\bf
x}},{\hat{\bf y}}$ plane.  When $\Phi$ is nonzero, it will refer to
the rotation of ${\hat{\bf a}}_1$ around ${\hat{\bf x}}$: e.g.,
$\Phi=90^\circ$ puts ${\hat{\bf a}}_1$ in the ${\hat{\bf x}},{\hat{\bf
z}}$ plane.

When $\beta=0$, ${\hat{\bf a}}_2$ will lie in the ${\hat{\bf
x}},{\hat{\bf a}}_1$ plane, in such a way that when $\Theta=0$ and
$\Phi=0$, ${\hat{\bf a}}_2$ is in the ${\hat{\bf y}}$ direction: e.g,
$\Theta=90^\circ$, $\Phi=0$, $\beta=0$ has ${\hat{\bf a}}_1={\hat{\bf
y}}$ and ${\hat{\bf a}}_2=-{\hat{\bf x}}$.  Nonzero $\beta$ introduces
an additional rotation of ${\hat{\bf a}}_2$ around ${\hat{\bf a}}_1$:
e.g., $\Theta=90^\circ$, $\Phi=0$, $\beta=90^\circ$ has ${\hat{\bf
a}}_1={\hat{\bf y}}$ and ${\hat{\bf a}}_2={\hat{\bf z}}$.

Mathematically:
\begin{eqnarray}
{\hat{\bf a}}_1 &=&   {\hat{\bf x}} \cos\Theta 
+ {\hat{\bf y}} \sin\Theta \cos\Phi 
+ {\hat{\bf z}} \sin\Theta \sin\Phi
	\\
{\hat{\bf a}}_2 &=& - {\hat{\bf x}} \sin\Theta \cos\beta 
+ {\hat{\bf y}} [\cos\Theta \cos\beta \cos\Phi-\sin\beta \sin\Phi] \nonumber\\
&&+ {\hat{\bf z}} [\cos\Theta \cos\beta \sin\Phi+\sin\beta \cos\Phi]
	\\
{\hat{\bf a}}_3 &=&   {\hat{\bf x}} \sin\Theta \sin\beta 
- {\hat{\bf y}} [\cos\Theta \sin\beta \cos\Phi+\cos\beta \sin\Phi] \nonumber\\
  &&         - {\hat{\bf z}} [\cos\Theta \sin\beta \sin\Phi-\cos\beta \cos\Phi]
\end{eqnarray}
or, equivalently:
\begin{eqnarray}
{\hat{\bf x}} &=&   {\hat{\bf a}}_1 \cos\Theta
           - {\hat{\bf a}}_2 \sin\Theta \cos\beta
           + {\hat{\bf a}}_3 \sin\Theta \sin\beta \\
{\hat{\bf y}} &=&   {\hat{\bf a}}_1 \sin\Theta \cos\Phi
           + {\hat{\bf a}}_2 [\cos\Theta \cos\beta \cos\Phi-\sin\beta \sin\Phi]
\nonumber\\
&&           - {\hat{\bf a}}_3 [\cos\Theta \sin\beta \cos\Phi+\cos\beta \sin\Phi]
\\
{\hat{\bf z}} &=&   {\hat{\bf a}}_1 \sin\Theta \sin\Phi
           + {\hat{\bf a}}_2 [\cos\Theta \cos\beta \sin\Phi+\sin\beta \cos\Phi]
\nonumber\\
&&           - {\hat{\bf a}}_3 [\cos\Theta \sin\beta \sin\Phi-\cos\beta \cos\Phi]
\end{eqnarray}

\subsection{Orientation of the Incident Beam in the Target Frame}

Under some circumstances, one may wish to specify the target
orientation such that ${\hat{\bf x}}$ (the direction of propagation of
the radiation) and ${\hat{\bf y}}$ (usually the first polarization
direction) and ${\hat{\bf z}}$ (= ${\hat{\bf x}} \times {\hat{\bf
y}})$ refer to certain directions in the TF.  Given the definitions of
the LF and TF above, this is simply an exercise in coordinate
transformation.  For example, one might wish to have the incident
radiation propagating along the (1,1,1) direction in the TF (example
14 below).  Here we provide some selected examples:
\begin{enumerate}
\item${\hat{\bf x}}= {\hat{\bf a}}_1$, ${\hat{\bf y}}= {\hat{\bf a}}_2$, ${\hat{\bf z}}= {\hat{\bf a}}_3$ : 
	$\Theta=  0$, $\Phi+\beta=  0$
\item${\hat{\bf x}}= {\hat{\bf a}}_1$, ${\hat{\bf y}}= {\hat{\bf a}}_3$, ${\hat{\bf z}}=-{\hat{\bf a}}_2$ : 
	$\Theta=  0$, $\Phi+\beta= 90^\circ$
\item${\hat{\bf x}}= {\hat{\bf a}}_2$, ${\hat{\bf y}}= {\hat{\bf a}}_1$, ${\hat{\bf z}}=-{\hat{\bf a}}_3$ : 
	$\Theta= 90^\circ$, $\beta=180^\circ$, $\Phi=  0$
\item${\hat{\bf x}}= {\hat{\bf a}}_2$, ${\hat{\bf y}}= {\hat{\bf a}}_3$, ${\hat{\bf z}}= {\hat{\bf a}}_1$ : 
	$\Theta= 90^\circ$, $\beta=180^\circ$, $\Phi= 90^\circ$
\item${\hat{\bf x}}= {\hat{\bf a}}_3$, ${\hat{\bf y}}= {\hat{\bf a}}_1$, ${\hat{\bf z}}= {\hat{\bf a}}_2$ : 
	$\Theta= 90^\circ$, $\beta=-90^\circ$, $\Phi=  0$
\item${\hat{\bf x}}= {\hat{\bf a}}_3$, ${\hat{\bf y}}= {\hat{\bf a}}_2$, ${\hat{\bf z}}=-{\hat{\bf a}}_1$ : 
	$\Theta= 90^\circ$, $\beta=-90^\circ$, $\Phi=-90^\circ$
\item${\hat{\bf x}}=-{\hat{\bf a}}_1$, ${\hat{\bf y}}= {\hat{\bf a}}_2$, ${\hat{\bf z}}=-{\hat{\bf a}}_3$ : 
	$\Theta=180^\circ$, $\beta+\Phi=180^\circ$
\item${\hat{\bf x}}=-{\hat{\bf a}}_1$, ${\hat{\bf y}}= {\hat{\bf a}}_3$, ${\hat{\bf z}}= {\hat{\bf a}}_2$ : 
	$\Theta=180^\circ$, $\beta+\Phi= 90^\circ$
\item${\hat{\bf x}}=-{\hat{\bf a}}_2$, ${\hat{\bf y}}= {\hat{\bf a}}_1$, ${\hat{\bf z}}= {\hat{\bf a}}_3$ : 
	$\Theta= 90^\circ$, $\beta=  0$, $\Phi=  0$
\item${\hat{\bf x}}=-{\hat{\bf a}}_2$, ${\hat{\bf y}}= {\hat{\bf a}}_3$, ${\hat{\bf z}}=-{\hat{\bf a}}_1$ : 
	$\Theta= 90^\circ$, $\beta=  0$, $\Phi=-90^\circ$
\item${\hat{\bf x}}=-{\hat{\bf a}}_3$, ${\hat{\bf y}}= {\hat{\bf a}}_1$, ${\hat{\bf z}}=-{\hat{\bf a}}_2$ : 
	$\Theta= 90^\circ$, $\beta=-90^\circ$, $\Phi=  0$
\item${\hat{\bf x}}=-{\hat{\bf a}}_3$, ${\hat{\bf y}}= {\hat{\bf a}}_2$, ${\hat{\bf z}}= {\hat{\bf a}}_1$ : 
	$\Theta= 90^\circ$, $\beta=-90^\circ$, $\Phi= 90^\circ$
\item${\hat{\bf x}}=({\hat{\bf a}}_1+{\hat{\bf a}}_2)/\surd2$, ${\hat{\bf y}}={\hat{\bf a}}_3$, 
	${\hat{\bf z}}=({\hat{\bf a}}_1-{\hat{\bf a}}_2)/\surd2$: 
	$\Theta=45^\circ$, $\beta=180^\circ$, $\Phi=90^\circ$
\item${\hat{\bf x}}=({\hat{\bf a}}_1+{\hat{\bf a}}_2+{\hat{\bf a}}_3)/\surd3$, 
	${\hat{\bf y}}=({\hat{\bf a}}_1-{\hat{\bf a}}_2)/\surd2$, 
	${\hat{\bf z}}=({\hat{\bf a}}_1+{\hat{\bf a}}_2-2{\hat{\bf a}}_3)/\surd6$:
	$\Theta=54.7356^\circ$, $\beta=135^\circ$, $\Phi=30^\circ$.
\end{enumerate}

\subsection{Sampling in $\Theta$, $\Phi$, and $\beta$\label{subsec:sampling}}

The present version, \ddscat, chooses the angles $\beta$,
$\Theta$, and $\Phi$ to sample the intervals ({\tt BETAMI,BETAMX}),
({\tt THETMI,THETMX)}, ({\tt PHIMIN,PHIMAX}), where {\tt BETAMI}, {\tt
BETAMX}, {\tt THETMI}, {\tt THETMX}, {\tt PHIMIN}, {\tt PHIMAX} are
specified in {\tt ddscat.par} .  The prescription for choosing the
angles is to:
\begin{itemize}
\item uniformly sample in $\beta$;
\item uniformly sample in $\Phi$;
\item uniformly sample in $\cos\Theta$.
\end{itemize}
This prescription is appropriate for random orientation of the target,
within the specified limits of $\beta$, $\Phi$, and $\Theta$.

Note that when \ddscat\ chooses angles it handles $\beta$ and
$\Phi$ differently from $\Theta$.\footnote{
	This is a change from {{\bf DDSCAT}}{\bf .4a}!!.
	}
The range for $\beta$ is divided into {\tt NBETA} intervals, and the
midpoint of each interval is taken.  Thus, if you take {\tt
BETAMI}=0, {\tt BETAMX}=90, {\tt NBETA}=2 you will get
$\beta=22.5^\circ$ and $67.5^\circ$.  Similarly, if you take
{\tt PHIMIN}=0, {\tt PHIMAX}=180, {\tt NPHI}=2 you will get
$\Phi=45^\circ$ and $135^\circ$.

Sampling in $\Theta$ is done quite differently from sampling in
$\beta$ and $\Phi$.  First, as already mentioned above, {{\bf
DDSCAT.6.0}}\ samples uniformly in $\cos\Theta$, not $\Theta$.
Secondly, the sampling depends on whether {\tt NTHETA} is even or odd.
\begin{itemize}
\item If {\tt NTHETA} is odd, then the values of $\Theta$ selected
	include the extreme values {\tt THETMI} and {\tt THETMX}; thus, {\tt
	THETMI}=0, {\tt THETMX}=90, {\tt NTHETA}=3 will give you
	$\Theta=0,60^\circ,90^\circ$.
\item If {\tt NTHETA} is even, then the range of $\cos\Theta$ will be
	divided into {\tt NTHETA} intervals, and the midpoint of each interval
	will be taken; thus, {\tt THETMI}=0, {\tt THETMX}=90, {\tt NTHETA}=2
	will give you $\Theta=41.41^\circ$ and $75.52^\circ$
	[$\cos\Theta=0.25$ and $0.75$].
\end{itemize}
The reason for this is that if odd {\tt NTHETA} is specified, than the
``integration'' over $\cos\Theta$ is performed using Simpson's rule
for greater accuracy.  If even {\tt NTHETA} is specified, then the
integration over $\cos\Theta$ is performed by simply taking the
average of the results for the different $\Theta$ values.

If averaging over orientations is desired, it is recommended that the
user specify an {\it odd} value of {\tt NTHETA} so that Simpson's rule
will be employed.

\section{Orientational Averaging\label{sec:orientational_averaging}}

{{\bf DDSCAT}} has been constructed to facilitate the computation of
orientational averages.  How to go about this depends on the
distribution of orientations which is applicable.

\subsection{Randomly-Oriented Targets}

For randomly-oriented targets, we wish to compute the orientational
average of a quantity $Q(\beta,\Theta,\Phi)$:
\begin{equation}
\langle Q \rangle = {1\over 8\pi^2}\int_0^{2\pi}d\beta
\int_{-1}^1 d\cos\Theta
\int_0^{2\pi}d\Phi ~ Q(\beta,\Theta,\Phi) ~~~.
\end{equation}
To compute such averages, all you need to do is edit the file {\tt
ddscat.par} so that DDSCAT knows what ranges of the angles $\beta$,
$\Theta$, and $\Phi$ are of interest.  For a randomly-oriented target
with no symmetry, you would need to let $\beta$ run from 0 to
$360^\circ$, $\Theta$ from 0 to $180^\circ$, and $\Phi$ from 0 to
$360^\circ$.

For targets with symmetry, on the other hand, the ranges of $\beta$,
$\Theta$, and $\Phi$ may be reduced.  First of all, remember that
averaging over $\Phi$ is relatively ``inexpensive", so when in doubt
average over 0 to $360^\circ$; most of the computational ``cost" is
associated with the number of different values of ($\beta$,$\Theta$)
which are used.  Consider a cube, for example, with axis ${\hat{\bf
a}}_1$ normal to one of the cube faces; for this cube $\beta$ need run
only from 0 to $90^\circ$, since the cube has fourfold symmetry for
rotations around the axis ${\hat{\bf a}}_1$.  Furthermore, the angle
$\Theta$ need run only from 0 to $90^\circ$, since the orientation
($\beta$,$\Theta$,$\Phi$) is indistinguishable from ($\beta$,
$180^\circ-\Theta$, $360^\circ-\Phi$).

For targets with symmetry, the user is encouraged to test the
significance of $\beta$,$\Theta$,$\Phi$ on targets with small numbers
of dipoles (say, of the order of 100 or so) but having the desired
symmetry.

It is important to remember that {{\bf DDSCAT}}{\bf .4b} handled even
and odd values of {\tt NTHETA} differently -- see
\S\ref{sec:parameter_file} above!  For averaging over random
orientations odd values of {\tt NTHETA} are to be preferred, as this
will allow use of Simpson's rule in evaluating the ``integral" over
$\cos\Theta$.

\subsection{Nonrandomly-Oriented Targets}

Some special cases (where the target orientation distribution is
uniform for rotations around the $x$ axis = direction of propagation
of the incident radiation), one may be able to use \ddscat\ 
with appropriate choices of input parameters.  More generally,
however, you will need to modify subroutine {\tt ORIENT} to generate a
list of {\tt NBETA} values of $\beta$, {\tt NTHETA} values of
$\Theta$, and {\tt NPHI} values of $\Phi$, plus two weighting arrays
{\tt WGTA(1-NTHETA,1-NPHI)} and {\tt WGTB(1-NBETA)}.  Here {\tt WGTA}
gives the weights which should be attached to each ($\Theta$,$\Phi$)
orientation, and {\tt WGTB} gives the weight to be attached to each
$\beta$ orientation.  Thus each orientation of the target is to be
weighted by the factor {\tt WGTA}$\times${\tt WGTB}.  For the case of
random orientations, \ddscat\ chooses $\Theta$ values which
are uniformly spaced in $\cos\Theta$, and $\beta$ and $\Phi$ values
which are uniformly spaced, and therefore uses uniform
weights\hfill\break \indent\indent {\tt WGTB}=1./{\tt
NBETA}\hfill\break 

\noindent When {\tt NTHETA} is even, {{\bf DDSCAT}}\ sets\hfill\break
\indent\indent {\tt WGTA}=1./({\tt NTHETA}$\times${\tt
NPHI})\hfill\break 

\noindent but when {\tt NTHETA} is odd, {{\bf DDSCAT}}\ uses Simpson's
rule when integrating over $\Theta$ and \hfill\break \indent\indent
{\tt WGTA}= (1/3 or 4/3 or 2/3)/({\tt NTHETA}$\times${\tt NPHI})

Note that the program structure of {{\bf DDSCAT}}\ may not be ideally
suited for certain highly oriented cases.  If, for example, the
orientation is such that for a given $\Phi$ value only one $\Theta$
value is possible (this situation might describe ice needles oriented
with the long axis perpendicular to the vertical in the Earth's
atmosphere, illuminated by the Sun at other than the zenith) then it
is foolish to consider all the combinations of $\Theta$ and $\Phi$
which the present version of {{\bf DDSCAT}}\ is set up to do.  We hope
to improve this in a future version of {{\bf DDSCAT}}.

\section{Target Generation\label{sec:target_generation}}

DDSCAT contains routines to generate dipole arrays representing
targets of various geometries, including spheres, ellipsoids,
rectangular solids, cylinders, hexagonal prisms, tetrahedra, two
touching ellipsoids, and three touching ellipsoids.  The target type
is specified by variable {\tt CSHAPE} on line 9 of {\tt ddscat.par},
up to 6 target shape parameters ({\tt SHPAR1}, {\tt SHPAR2}, {\tt
SHPAR3}, ...) on line 10, and the lattice spacing {\tt DX DY DZ} on
line 11 (for a cubic lattice, set {DX DY DZ} to {\tt 1.\ 1.\ 1.} on line
11).  The target geometry is most conveniently described in a
coordinate system attached to the target which we refer to as the
``Target Frame'' (TF), so in this section {\it only} we will let x,y,z
be coordinates in the Target Frame.  Once the target is generated, the
orientation of the target in the Lab Frame is accomplished as
described in \S\ref{sec:target_orientation}.

Target geometries currently supported include:\footnote{
	In DDSCAT.6.0 not all of the target generation routines
	have been modified for use with a noncubic lattice.
	Those routines which have not yet been modified
	({\tt TAR2EL,TAR2SP,TAR3EL,TARCEL,TARHEX,TARTET} 
	include code to ensure that they are not
	used with a noncubic lattice; if called with a noncubic
	lattice, an error message is generated 
	(e.g. ``tarhex does not support noncubic lattice'')
	and execution is terminated.
	}
\begin{itemize}
\item {\tt ELLIPS} -- {\bf Homogeneous, isotropic ellipsoid.}\\
	``Lengths'' 
	{\tt SHPAR1}, {\tt SHPAR2},
	{\tt SHPAR3} in the $x$, $y$, $z$ directions in the TF.
	$(x/{\tt SHPAR1})^2+(y/{\tt SHPAR2})^2+(z/{\tt SHPAR3})^2 = d^2/4$,
	where $d$ is the interdipole spacing.\\
	The target axes are set to ${\hat{\bf a}}_1=(1,0,0)$ and 
	${\hat{\bf a}}_2=(0,1,0)$ in the TF.\\
	User must set {\tt NCOMP}=1 on line 9 of {\tt ddscat.par}.\\
	A {\bf homogeneous, isotropic sphere} is obtained by setting 
	{\tt SHPAR1} = {\tt SHPAR2} = {\tt SHPAR3} = diameter/$d$.
\item {\tt ANIELL} -- {\bf Homogeneous, anisotropic ellipsoid.}\\
	{\tt SHPAR1}, {\tt SHPAR2},
	{\tt SHPAR3} have same meaning as for {\tt ELLIPS}.
	Target axes ${\hat{\bf a}}_1=(1,0,0)$ and 
	${\hat{\bf a}}_2=(0,1,0)$ in the TF. 
	It is assumed that the dielectric tensor is diagonal in the TF. 
	User must set {\tt NCOMP}=3 and provide $xx$, $yy$, $zz$ elements of
	the dielectric tensor.
\item {\tt CONELL} -- {\bf Two concentric ellipsoids.}\\
	{\tt SHPAR1}, {\tt SHPAR2},
	{\tt SHPAR3} specify the lengths along $x$, $y$, $z$ axes (in the TF)
	of the {\it outer} ellipsoid; {\tt SHPAR4}, {\tt SHPAR5}, {\tt SHPAR6} 
	are the lengths, along the $x$, $y$, $z$ axes (in the TF) of the
	{\it inner} ellipsoid.  
	The ``core" within the inner ellipsoid is composed of isotropic 
	material 1; 
	the ``mantle" between inner and outer
	ellipsoids is composed of isotropic material 2.  
	Target axes ${\hat{\bf a}}_1=(1,0,0)$, ${\hat{\bf a}}_2=(0,1,0)$ 
	in TF.  
	User must set {\tt NCOMP}=2 and provide dielectric functions for 
	``core'' and ``mantle'' materials.
\item {\tt CYLNDR} -- {\bf Homogeneous, isotropic cylinder.}\\
	Length/$d$={\tt SHPAR1}, 
	diameter/$d$={\tt SHPAR2}, with cylinder axis 
	$={\hat{\bf a}}_1=(1,0,0)$
	and ${\hat{\bf a}}_2=(0,1,0)$ in the TF.
	User must set {\tt NCOMP}=1.
\item {\tt UNICYL} -- {\bf Homogeneous cylinder with unixial anisotropic 
	dielectric tensor.}\\
	{\tt SHPAR1}, {\tt SHPAR2} have same meaning as for
	{\tt CYLNDR}.
	Cylinder axis $={\hat{\bf a}}_1=(1,0,0)$, ${\hat{\bf a}}_2=(0,1,0)$. 
	It is assumed that the dielectric tensor $\epsilon$ 
	is diagonal in the TF,
	with $\epsilon_{yy}=\epsilon_{zz}$.
	User must set
	{\tt NCOMP}=2.
	Dielectric function 1 is for ${\bf E} \parallel {\bf\hat{a}}_1$
	(cylinder axis), dielectric function 2 is for 
	${\bf E} \perp {\bf\hat{a}}_1$.
\item {\tt RCTNGL} -- {\bf Homogeneous, isotropic, rectangular solid.}\\
	x, y, z lengths/$d$ = {\tt SHPAR1}, {\tt SHPAR2}, {\tt SHPAR3}.  
	Target axes ${\hat{\bf a}}_1=(1,0,0)$ and ${\hat{\bf a}}_2=(0,1,0)$ 
	in the TF.
	User must set {\tt NCOMP}=1.
\item {\tt HEXGON} -- {\bf Homogeneous, isotropic hexagonal prism.}\\
	Length/$d$ = 
	{\tt SHPAR1}, 
	hexagon side/$d$ = {\tt SHPAR2}.
	Target axis ${\hat{\bf a}}_1 = (1,0,0)$ in the TF is along the
	prism axis, and target
	axis ${\hat{\bf a}}_2=(0,1,0)$ in the TF is
	normal to one of the rectangular faces
	of the hexagonal prism.
	User must set {\tt NCOMP}=1.
\item {\tt TETRAH} -- {\bf Homogeneous, isotropic tetrahedron.}\\
	{\tt SHPAR1}=length/$d$ 
	of one edge. 
	Orientation: one face parallel to $y$,$z$ plane (in the TF), 
	opposite ``vertex" is in $+x$ 
	direction, and one edge is parallel to $z$ axis (in the TF).  
	Target axes ${\hat{\bf a}}_1=(1,0,0)$ [emerging from one vertex] and 
	${\hat{\bf a}}_2=(0,1,0)$ [emerging from an edge] in the TF.
	User must set {\tt NCOMP}=1.
\item {\tt TWOSPH} -- {\bf Two touching homogeneous, isotropic spheroids,
	with distinct compositions.}\\ 
	First spheroid has length {\tt SHPAR1} along symmetry axis, diameter 
	{\tt SHPAR2} perpendicular to symmetry axis.
	Second spheroid has length {\tt SHPAR3} along symmetry axis, 
	diameter {\tt SHPAR4} perpendicular to symmetry axis.  
	Contact point is on line connecting centroids.  
	Line connecting centroids is in $x$ direction.
	Symmetry axis of first spheroid is in $y$ direction.  
	Symmetry axis of second spheroid is in direction 
	${\hat{\bf y}}\cos({\tt SHPAR5})+{\hat{\bf z}}\sin({\tt SHPAR5})$,
	where ${\hat{\bf y}}$ and ${\hat{\bf z}}$ are basis vectors in TF, 
	and {\tt SHPAR5} is in degrees.  
	If {\tt SHPAR6}=0., then target axes ${\hat{\bf a}}_1=(1,0,0)$,
	${\hat{\bf a}}_2=(0,1,0)$. 
	If {\tt SHPAR6}=1., then axes ${\hat{\bf a}}_1$ and ${\hat{\bf a}}_2$ 
	are set to 
	principal axes with largest and 2nd largest moments of inertia assuming
	spheroids to be of uniform density.
	User must set {\tt NCOMP}=2 and provide dielectric function files for 
	each spheroid.
\item {\tt TWOELL} -- {\bf Two touching, homogeneous, isotropic ellipsoids,
	with distinct compositions.}\\
	{\tt SHPAR1}, {\tt SHPAR2}, {\tt SHPAR3}=x-length/$d$, y-length/$d$,
	$z$-length/$d$ of one ellipsoid.
	The two ellipsoids have identical shape, size, and orientation,
	but distinct dielectric functions.
	The line connecting ellipsoid centers is along the $x$-axis in the TF.
	Target axes ${\hat{\bf a}}_1=(1,0,0)$ 
	[along line connecting ellipsoids]
	and ${\hat{\bf a}}_2=(0,1,0)$.
	User must set {\tt NCOMP}=2 and provide dielectric function file names
	for both ellipsoids. 
	Ellipsoids are in order of increasing x:
	first dielectric function is for ellipsoid with 
	center at negative $x$, second dielectric function for 
	ellipsoid with center at positive $x$.
\item {\tt TWOAEL} -- {\bf Two touching, homogeneous, anisotropic 
	ellipsoids, with distinct compositions.}\\ 
	Geometry as for {\tt TWOELL};
	{\tt SHPAR1}, {\tt SHPAR2}, {\tt SHPAR3} have same meanings as for 
	{\tt TWOELL}.  
	Target axes ${\hat{\bf a}}_1=(1,0,0)$ and ${\hat{\bf a}}_2=(0,1,0)$
	in the TF.
	It is assumed that (for both ellipsoids) the dielectric tensor
	is diagonal in the TF.
	User must set {\tt NCOMP}=6 and provide 
	$xx$, $yy$, $zz$ components of dielectric tensor for 
	first ellipsoid, and $xx$, $yy$, $zz$ components of 
	dielectric tensor for second 
	ellipsoid (ellipsoids are in order of increasing x).
\item {\tt THRELL} -- {\bf Three touching homogeneous, isotropic ellipsoids 
	of equal size and orientation, but distinct compositions.}\\
	{\tt SHPAR1}, {\tt SHPAR2}, {\tt SHPAR3} have same meaning as for 
	{\tt TWOELL}.
	Line connecting ellipsoid centers is parallel to $x$ axis.  
	Target axis ${\hat{\bf a}}_1=(1,0,0)$ along line of ellipsoid centers, 
	${\hat{\bf a}}_2=(0,1,0)$.
	User must set {\tt NCOMP}=3 and provide (isotropic) dielectric 
	functions for first, second, and third ellipsoid.
\item {\tt THRAEL} -- {\bf Three touching homogeneous, anisotropic ellipsoids 
	with same size and 
	orientation but distinct dielectric tensors.}\\
	{\tt SHPAR1}, {\tt SHPAR2}, {\tt SHPAR3} have same meanings as for 
	{\tt THRELL}.  
	Target axis ${\hat{\bf a}}_1=(1,0,0)$ along line of ellipsoid centers, 
	${\hat{\bf a}}_2=(0,1,0)$.  
	It is assumed that dielectric tensors 
	are all diagonal in the TF.
	User must set {\tt NCOMP}=9 and provide $xx$, $yy$, $zz$ 
	elements of dielectric tensor
	for first ellipsoid, $xx$, $yy$, $zz$ elements for second ellipsoid, 
	and $xx$, $yy$, $zz$ elements for third ellipsoid 
	(ellipsoids are in order of increasing $x$).
\item {\tt BLOCKS} -- {\bf Homogeneous target constructed from 
	cubic ``blocks''.}\\  
	Number and location of blocks are specified in separate file 
	{\tt blocks.par} with following structure:\\
	\hspace*{2em}one line of comments (may be blank)\\
	\hspace*{2em}{\tt PRIN} (= 0 or 1 -- see below)\\
	\hspace*{2em}{\tt N} (= number of blocks) \\
	\hspace*{2em}{\tt B}  (= width/$d$ of one block) \\
	\hspace*{2em}$x$ $y$ $z$ (= position of 1st block 
		in units of {\tt B}$d$)\\
	\hspace*{2em}$x$ $y$ $z$ (= position of 2nd block)\\
	\hspace*{2em}... \\
	\hspace*{2em}$x$ $y$ $z$ (= position of {\tt N}th block)\\
	If {\tt PRIN}=0, then ${\hat{\bf a}}_1=(1,0,0)$, 
	${\hat{\bf a}}_2=(0,1,0)$.
	If {\tt PRIN}=1, then ${\hat{\bf a}}_1$ and ${\hat{\bf a}}_2$ are set 
	to principal 
	axes with largest and second largest moments of inertia,
	assuming target to be of uniform density.
	User must set {\tt NCOMP=1}.
\item {\tt DW1996} -- {\bf 13 block target used by 
	Draine \& Weingartner (1996).} \\
	Single, isotropic material.
	Target geometry was used in study by Draine \& Weingartner (1996) 
	of radiative torques on irregular grains.
	${\hat{\bf a}}_1$ and ${\hat{\bf a}}_2$ 
	are principal axes with largest and
	second-largest moments of inertia.
	User must set {\tt NCOMP=1}.
	Target size is controlled by shape parameter {\tt SHPAR(1)} = 
	width of one block in lattice units.
\item {\tt NSPHER} -- {\bf Multisphere target consisting of the union of $N$
	spheres of single isotropic material.}\\  
	Spheres may overlap if desired.
	The relative locations and sizes of these spheres are
	defined in an external file, whose name (enclosed in quotes) 
	is passed through {\tt ddscat.par}.  The length of the file name
	should not exceed 13 characters.  
	The pertinent line in {\tt ddscat.par} should read\\
	{\tt SHPAR(1) SHPAR(2)} {\it `filename'} (quotes must be used)\\
	where {\tt SHPAR(1)} = target diameter in $x$ direction 
	(in Target Frame) in units of $d$\\
	{\tt SHPAR(2)}= 0 to have $a_1=(1,0,0)$, $a_2=(0,1,0)$ 
	in Target Frame.\\
	{\tt SHPAR(2)}= 1 to use principal axes of moment of inertia
	tensor for $a_1$ (largest $I$) and $a_2$ (intermediate $I$).\\
	{\it filename} is the name of the file specifying the locations and
	relative sizes of the spheres.\\
	The overall size of the multisphere target (in terms of numbers of
	dipoles) is determined by parameter {\tt SHPAR(1)}, which is
	the extent of the multisphere target in the $x$-direction, in
	units of the lattice spacing $d$.
	The file {\it `filename'} should have the
	following structure:\\
	\hspace*{2em}$N$ (= number of spheres)\\
	\hspace*{2em}one line of comments (may be blank)\\
	\hspace*{2em}$x_1$ $y_1$ $z_1$ $a_1$ (arb. units)\\
	\hspace*{2em}$x_2$ $y_2$ $z_2$ $a_2$ (arb. units)\\
	\hspace*{2em} ... \\
	\hspace*{2em}$x_N$ $y_N$ $z_N$ $a_N$ (arb. units)\\
	where $x_j$, $y_j$, $z_j$ are the coordinates of the center,
	and $a_j$ is the radius of sphere $j$.\\
	Note that $x_j$, $y_j$, $z_j$, $a_j$ ($j=1,...,N$) establish only
	the {\it shape} of the $N-$sphere target.  For instance,
	a target consisting of two touching spheres with the line between
	centers parallel to the $x$ axis could equally well be
	described by lines 3 and 4 being\\
	\hspace*{2em}0 0 0 0.5\\
	\hspace*{2em}1 0 0 0.5\\
	or\\
	\hspace*{2em}0 0 0 1\\
	\hspace*{2em}2 0 0 1\\
	The actual size (in physical units) is set by the value
	of $a_{\rm eff}$ specified in {\tt ddscat.par}, where, as
	always, $a_{\rm eff}\equiv (3 V/4\pi)^{1/3}$, where $V$ is the
	volume of material in the target.\\
	User must set {\tt NCOMP}=1.
\item {\tt PRISM3} -- {\bf Triangular prism of 
	homogeneous, isotropic material.}\\
	{\tt SHPAR1, SHPAR2, SHPAR3, SHPAR4} $= a/d$, $b/a$, $c/a$, $L/a$\\
	The triangular cross section has sides of width $a$, $b$, $c$.
	$L$ is the length of the prism.
	$d$ is the lattice spacing.
	The triangular cross-section 
	has interior angles $\alpha$, $\beta$, $\gamma$
	(opposite sides $a$, $b$, $c$) given by
	$\cos\alpha=(b^2+c^2-a^2)/2bc$, $\cos\beta=(a^2+c^2-b^2)/2ac$,
	$\cos\gamma=(a^2+b^2-c^2)/2ab$.
	In the Target Frame, the prism axis is in the $\hat{\bf x}$ direction,
	the normal to the rectangular face of width $a$ is (0,1,0), 
	the normal to the rectangular face of width $b$ is
	$(0,-\cos\gamma,\sin\gamma)$, and
	the normal to the rectangular face of width  $c$ is
	$(0,-\cos\gamma,-\sin\gamma)$.

\item {\tt LYRSLB} -- Multilayer rectangular slab with overall x, y, z
	lengths 
	$a_x$ = {\tt SHPAR1}$\times d$, 
	$a_y$ = {\tt SHPAR2}$\times d$, 
	$a_z$ = {\tt SHPAR3}$\times d$,
	with the boundary between layers 1 and 2 occuring at
	$x_1={\tt SHPAR4}\times a_x$,
	boundary between layers 2 and 3 occuring at
	$x_2={\tt SHPAR5}\times a_x$ [if {\tt SHPAR5}$>$ 0],
	and boundary between layers 3 and 4 occuring at
	$x_3={\tt SHPAR6}\times a_x$ [if {\tt SHPAR6}$>$ 0],
	To create a simple bilayer slab, just set 
	{\tt SHPAR5}={\tt SHPAR6}=0.
	User must set {\tt NCOMP}=2,3, or 4 and provide dielectric function
	files for each of the two layers.

\item {\tt FRMFIL} -- {\bf List of dipole locations and
	``compositions'' obtained from a file.}\\
	This option causes {{\bf DDSCAT}}\ to read the target geometry
	information from a file {\tt shape.dat} instead of automatically
	generating one of the geometries listed above.  The {\tt shape.dat}
	file is read by routine {\tt REASHP} (file {\tt reashp.f}).
	The user can customize {\tt REASHP} as needed to conform to the
	manner in which the target geometry is stored in file {\tt shape.dat}.
	However, as supplied, {\tt REASHP} expects the file {\tt shape.dat}
	to have the following structure:
	\begin{itemize}
	\item one line containing a description; the first 67 characters will
		be read and printed in various output statements.
	\item {\tt N} = number of dipoles in target
	\item $a_{1x}$ $a_{1y}$ $a_{1z}$ = x,y,z components of $\bf{a}_1$
	\item $a_{2x}$ $a_{2y}$ $a_{2z}$ = x,y,z components of $\bf{a}_2$
	\item (line containing comments)
	\item $dummy$ {\tt IXYZ(1,1) IXYZ(1,2) IXYZ(1,3) 
		ICOMP(1,1) ICOMP(1,2) ICOMP(1,3)}
	\item $dummy$ {\tt IXYZ(2,1) IXYZ(2,2) IXYZ(2,3) 
		ICOMP(2,1) ICOMP(2,2) ICOMP(2,3)}
	\item $dummy$ {\tt IXYZ(3,1) IXYZ(3,2) IXYZ(3,3)
		ICOMP(3,1) ICOMP(3,2) ICOMP(3,3)}
	\item ...
	\item $dummy$ {\tt IXYZ(J,1) IXYZ(J,2) IXYZ(J,3)
		ICOMP(J,1) ICOMP(J,2) ICOMP(J,3)}
	\item ...
	\item $dummy$ {\tt IXYZ(N,1) IXYZ(N,2) IXYZ(N,3) 
		ICOMP(N,1) ICOMP(N,2) ICOMP(N,3)}
	\end{itemize}
\end{itemize}

The user should be able to easily modify these routines, or write new
routines, to generate targets with other geometries.  The user should
first examine the routine {\tt target.f} and modify it to call any new
target generation routines desired.  Alternatively, targets may be
generated separately, and the target description (locations of dipoles
and ``composition" corresponding to x,y,z dielectric properties at
each dipole site) read in from a file by invoking the option {\tt
FRMFIL} in {\tt ddscat.f}.

It is often desirable to be able to run the target generation routines
without running the entire {{\bf DDSCAT}}\ code.  We have therefore
provided a program {\tt CALLTARGET} which allows the user to generate
targets interactively; to create this executable just type\footnote{
	Non-Unix sites: The source code for {\tt CALLTARGET} is in the file
	{\tt MISC.FOR}.  You must compile and link this to {\tt ERRMSG}, {\tt
	GETSET}, {\tt REASHP}, {\tt TAR2EL}, {\tt TAR2SP}, {\tt TAR3EL}, {\tt
	TARBLOCKS}, {\tt TARCEL}, {\tt TARCYL}, {\tt TARELL}, {\tt TARGET},
	{\tt TARHEX}, {\tt TARREC}, {\tt TARTET}, and {\tt WRIMSG}.  The
	source code for {\tt ERRMSG} is in {\tt SRC2.FOR}, {\tt GETSET} is in
	{\tt SRC5.FOR}, and the rest are in {\tt SRC8.FOR} and 
	{\tt SRC9.FOR} .
	}  
\hfill\break \indent\indent{\tt make calltarget} .\hfill\break 
The program {\tt calltarget} is to be run interactively; the prompts are
self-explanatory.  You may need to edit the code to change the device
number {\tt IDVOUT} as for {\tt DDSCAT} (see \S\ref{subsec:IDVOUT}
above).

After running, {\tt calltarget} will leave behind an ASCII file 
{\tt target.out}
which is a list of the occupied lattice sites in the last target generated.
The format of {\tt target.out} is the same as the format of the {\tt shape.dat}
files read if option {\tt FRMFIL} is used (see above).
Therefore you can simply \\
\indent\indent{\tt mv target.out shape.dat} \\
and then use {\tt DDSCAT}
with the option {\tt FRMFIL}
in order to input a target shape generated by {\tt CALLTARGET}.

\section{Scattering Directions\label{sec:scattering_directions}}

{{\bf DDSCAT}}\ calculates scattering in selected directions, and elements of
the scattering matrix are reported in the output 
files {\tt w}{\it xx}{\tt r}{\it yy}{\tt k}{\it zzz}{\tt .sca} .
The scattering direction is specified through angles $\theta_s$ and $\phi_s$ 
(not to be confused with the angles $\Theta$ and $\Phi$ which specify the 
orientation of the target relative to the incident radiation!).

The scattering angle $\theta_s$ is simply the angle between the incident beam
(along direction ${\hat{\bf x}}$) and the scattered beam ($\theta_s=0$ for 
forward scattering, $\theta_s=180^\circ$ for backscattering).

The scattering angle $\phi_s$ specifies the orientation of the ``scattering 
plane'' relative to the ${\hat{\bf x}},{\hat{\bf y}}$ plane in the Lab Frame.
When $\phi_s=0$ the scattering plane is assumed to coincide with the
${\hat{\bf x}},{\hat{\bf y}}$ plane.
When $\phi_s=90^\circ$ the scattering plane is assumed to coincide with
the ${\hat{\bf x}},{\hat{\bf z}}$ plane.
Within the scattering plane the scattering directions are specified by
$0\leq\theta_s\leq180^\circ$.

Scattering directions for which the scattering properties are to
be calculated are set in the file {\tt ddscat.par} by
specifying one or more scattering planes (determined by the value of $\phi_s$)
and for each scattering plane, the number and range of $\theta_s$ values.
The only limitation is that the number of scattering directions not exceed
the parameter {\tt MXSCA} in {\tt DDSCAT.f} (in the code as distributed it
is set to {\tt MXSCA=1000}).

\section{Incident Polarization State\label{sec:incident_polarization}}

Recall that the ``Lab Frame'' is defined such that the incident radiation
is propagating along the ${\hat{\bf x}}$ axis.
\ddscat\ allows the user to specify a 
general elliptical polarization
for the incident radiation, by specifying the (complex) polarization
vector ${\hat{\bf e}}_{01}$.
The orthonormal polarization state 
${\hat{\bf e}}_{02}={\hat{\bf x}}\times{\hat{\bf e}}_{01}^*$ is
generated automatically if {\tt ddscat.par} specifies {\tt IORTH=2}.

For incident linear polarization, one can simply set ${\hat{\bf
e}}_{01}={\hat{\bf y}}$ by specifying {\tt (0,0) (1,0) (0,0)} in {\tt
ddscat.par}; then ${\hat{\bf e}}_{02}={\hat{\bf z}}$ For polarization
mode ${\hat{\bf e}}_{01}$ to correspond to right-handed circular
polarization, set ${\hat{\bf e}}_{01}=({\hat{\bf y}}+i{\hat{\bf
z}})/\surd2$ by specifying {\tt (0,0) (1,0) (0,1)} in {\tt ddscat.par}
(\ddscat\ automatically takes care of the normalization of
${\hat{\bf e}}_{01}$); then 
${\hat{\bf e}}_{02}=(-i{\hat{\bf y}}+{\hat{\bf z}})/\surd2$, 
corresponding to left-handed circular polarization.

\section{Averaging over Scattering: $g(1)=\langle\cos\theta_s\rangle$, 
	etc.\label{sec:averaging_scattering}}

{{\bf DDSCAT}}\ automatically carries out numerical integration of various
scattering properties.  In particular, it calculates the mean value
$g(1)=\langle\cos\theta_s\rangle$ for the scattered intensity for each
incident polarization state.
This is accomplished by evaluating the scattered intensity for
{\tt ICTHM} different values of $\theta_s$ (including $\theta_s=0$
and $\theta_s=\pi$), and taking a weighted sum.
For each value of $\theta_s$ except $0$ and $\pi$, the scattering
intensity is evaluated for {\tt IPHM} different values of the scattering
angle $\phi_s$.
The angular integration over $\theta_s$ is accomplished using Simpson's
rule (with uniform intervals in $\cos\theta_s$), 
so {\tt ICTHM} should be an {\it odd} number.
The angular integration over $\phi_s$ is accomplished by sampling
uniformly in $\phi_s$ with uniform weighting; {\tt IPHM} can be either
even or odd.

The following quantities are evaluated by this angular integration:
\begin{itemize}
\item${\bf g}=\langle\cos\theta_s\rangle{\hat{\bf x}}+
	\langle\sin\theta_s\cos\phi_s\rangle{\hat{\bf y}}+
	\langle\sin\theta_s\sin\phi_s\rangle{\hat{\bf z}}$
	(see \S\ref{sec:torque_calculation});
\item${\bf Q}_{\Gamma}$ (see \S\ref{sec:torque_calculation}).
\end{itemize}
It is important that the user recognize that accurate evaluation 
of these angular averages requires adequate sampling over scattering angles.
For small values of the size parameter $x=2\pi a_{\rm eff}/\lambda$,
the angular distribution of scattered radiation has a dipolar character
and the sampling in $\theta_s$ and $\phi_s$ does not need to be very fine,
so {\tt ICTHM} and {\tt IPHM} need not be large.
For larger values of the size parameter $x$, however, higher multipoles
in the scattered radiation field become important, and finer sampling
in $\theta_s$ and $\phi_s$ is required.
We do not have any foolproof prescription to offer, since the scattering
pattern will depend upon the target geometry and dielectric constant
in addition to overall size parameter.
However, as a very rough guide, we suggest that the user specify
values of {\tt ICTHM} and {\tt IPHM} satisfying
\begin{eqnarray}
{\tt ICTHM} &>& 5(1+x)\label{eq:ICTHM}~~~,\\
{\tt IPHM} &>& 2(1+x) ~~~.
\label{eq:IPHM}
\end{eqnarray}
The sample {\tt ddscat.par} file supplied has ${\tt ICTHM=33}$ and
${\tt IPHM}=12$; the above criteria would suggest that this would
be suitable for $x<5$.

The cpu time required for evaluation of these angular averages
is proportional to $[2+{\tt IPHM}({\tt ICTHM}-2)]$.
Since the computational time spent in evaluating these angular
integrals can be a significant part of the total, it is important
to choose values of {\tt ICTHM} and {\tt IPHM} which will provide
a suitable balance between accuracy (in this part of the overall calculation)
and cpu time.

Within one scattering plane, the scattered intensity tends to
have approximately $(1+x)$ peaks for $0\leq\theta_s\leq\pi$, so that
the above prescription for {\tt ICTHM} would have at least 5 sampling
points per maximum.
The angular distribution over $\phi_s$ is usually not as
structured as that over $\theta_s$ so we suggest that {\tt IPHM} need not
be as large as {\tt ICTHM}.
We have refrained from ``hard-wiring'' the values of {\tt ICTHM}
and {\tt IPHM} because we are not confident of the reliability of the
recommended criteria (\ref{eq:ICTHM},\ref{eq:IPHM}) -- it is up to
the user to specify appropriate values of {\tt ICTHM} and {\tt IPHM}
according to the requirements of the problem being addressed.

\section{Mueller Matrix for Scattering in Selected Directions
\label{sec:mueller_matrix}}

\subsection{Two Orthogonal Incident Polarizations ({\tt IORTH=2})}

\ddscat\ internally computes the scattering properties of the
dipole array in terms of a complex scattering matrix
$f_{ml}(\theta_s,\phi_s)$ (Draine 1988), where index $l=1,2$ denotes
the incident polarization state, $m=1,2$ denotes the scattered
polarization state, and $\theta_s$,$\phi_s$ specify the scattering
direction.  Normally {{\bf DDSCAT}}\ is used with {\tt IORTH=2} in
{\tt ddscat.par}, so that the scattering problem will be solved for
both incident polarization states ($l=1$ and 2); in this subsection it
will be assumed that this is the case.

Incident polarization states $l=1,2$ correspond to polarization states
${\hat{\bf e}}_{01}$, ${\hat{\bf e}}_{02}$; recall that polarization
state ${\hat{\bf e}}_{01}$ is user-specified, and 
${\hat{\bf e}}_{02}=\hat{\bf x}\times{\hat{\bf e}}_{01}^*$.  Scattered
polarization state $m=1$ corresponds to linear polarization of the
scattered wave parallel to the scattering plane 
(${\hat{\bf e}}_1={\hat{\bf e}}_{\parallel s}=\hat{\theta}_s$) and $m=2$
corresponds to linear polarization perpendicular to the scattering
plane (in the $+\hat{\phi}_s$ direction).  
The scattering matrix $f_{ml}$ was defined (Draine 1988) so that 
the scattered electric field ${\bf E}_s$ is related to the incident 
electric field ${\bf E}_i(0)$ at the origin 
(where the target is assumed to be located) by
\begin{equation}
\left(
\begin{array}{c}
	{\bf E}_s\cdot\hat{\theta}_s\\
	{\bf E}_s\cdot\hat{\phi}_s
\end{array}
\right)
=
{\exp(i{\bf k}\cdot{\bf r})\over kr}
\left(
\begin{array}{cc}
	f_{11}&f_{12}\\
	f_{21}&f_{22}
\end{array}
\right)
\left(
\begin{array}{c}
	{\bf E}_i(0)\cdot{\hat{\bf e}}_{01}\\
	{\bf E}_i(0)\cdot{\hat{\bf e}}_{02}
\end{array}
\right) ~~~.
\label{eq:f_ml_def}
\end{equation}
The 2$\times$2 complex 
{\it amplitude scattering matrix} (with elements $S_1$, $S_2$,
$S_3$, and $S_4$) is defined so that (see Bohren \& Huffman 1983)
\begin{equation}
\left(
\begin{array}{c}
	{\bf E}_s\cdot\hat{\theta}_s\\
	-{\bf E}_s\cdot\hat{\phi}_s
\end{array}
\right)
=
{\exp(i{\bf k}\cdot{\bf r})\over -ikr}
\left(
\begin{array}{cc}
	S_2&S_3\\
	S_4&S_1
\end{array}
\right)
\left(
\begin{array}{c}
	{\bf E}_i(0)\cdot{\hat{\bf e}}_{i\parallel}\\
	{\bf E}_i(0)\cdot{\hat{\bf e}}_{i\perp}
\end{array}
\right)~~~,
\label{eq:S_ampl_def}
\end{equation}
where
${\hat{\bf e}}_{i\parallel}$, ${\hat{\bf e}}_{i\perp}$ are (real) unit vectors 
for incident polarization
parallel and perpendicular to the scattering plane (with the
customary definition of ${\hat{\bf e}}_{i\perp}={\hat{\bf e}}_{i\parallel}\times{\hat{\bf x}}$).

From (\ref{eq:f_ml_def},\ref{eq:S_ampl_def}) we may write
\begin{equation}
\left(
\begin{array}{cc}
	S_2&S_3\\
	S_4&S_1
\end{array}
\right)
\left(
\begin{array}{c}
	{\bf E}_i(0)\cdot{\hat{\bf e}}_{i\parallel}\\
	{\bf E}_i(0)\cdot{\hat{\bf e}}_{i\perp}
\end{array}
\right)
=
-i
\left(
\begin{array}{cc}
	f_{11}&f_{12}\\
	-f_{21}&-f_{22}
\end{array}
\right)
\left(
\begin{array}{c}
	{\bf E}_i(0)\cdot{\hat{\bf e}}_{01}^*\\
	{\bf E}_i(0)\cdot{\hat{\bf e}}_{02}^*
\end{array}
\right) ~~~.
\label{eq:fml_rel_S_v1}
\end{equation}

Let
\begin{eqnarray}
	a&\equiv& {\hat{\bf e}}_{01}^*\cdot{\hat{\bf y}}~~~,\\
	b&\equiv& {\hat{\bf e}}_{01}^*\cdot{\hat{\bf z}}~~~,\\
	c&\equiv& {\hat{\bf e}}_{02}^*\cdot{\hat{\bf y}}~~~,\\
	d&\equiv& {\hat{\bf e}}_{02}^*\cdot{\hat{\bf z}}~~~.
\end{eqnarray}
Note that since ${\hat{\bf e}}_{01}, {\hat{\bf e}}_{02}$ could be
complex (i.e., elliptical polarization),
the quantities $a,b,c,d$ are complex.
Then
\begin{equation}
\left(
\begin{array}{c}
	{\hat{\bf e}}_{01}^*\\
	{\hat{\bf e}}_{02}^*
\end{array}
\right)
=
\left(
\begin{array}{cc}
	a&b\\
	c&d
\end{array}
\right)
\left(
\begin{array}{c}
	\hat{\bf y}\\
	\hat{\bf z}
\end{array}
\right)
\end{equation}
and eq. (\ref{eq:fml_rel_S_v1}) can be written
\begin{equation}
\left(
\begin{array}{cc}
	S_2&S_3\\
	S_4&S_1
\end{array}
\right)
\left(
\begin{array}{c}
	{\bf E}_i(0)\cdot{\hat{\bf e}}_{i\parallel}\\
	{\bf E}_i(0)\cdot{\hat{\bf e}}_{i\perp}
\end{array}
\right)
=
i
\left(
\begin{array}{cc}
	-f_{11}&-f_{12}\\
	f_{21}&f_{22}
\end{array}
\right)
\left(
\begin{array}{cc}
	a&b\\
	c&d
\end{array}
\right)
\left(
\begin{array}{c}
	{\bf E}_i(0)\cdot\hat{\bf y}\\
	{\bf E}_i(0)\cdot\hat{\bf z}
\end{array}
\right) ~~~.
\label{eq:fml_rel_S_v2}
\end{equation}

The incident polarization states ${\hat{\bf e}}_{i\parallel}$ and
${\hat{\bf e}}_{i\perp}$ are related to $\hat{\bf y}$, $\hat{\bf z}$ by
\begin{equation}
\left(
\begin{array}{c}
	\hat{\bf y}\\ 
	\hat{\bf z}
\end{array}
\right)
=
\left(
\begin{array}{cc}
	\cos\phi_s&	\sin\phi_s\\
	\sin\phi_s&	-\cos\phi_s
\end{array}
\right)
\left(
\begin{array}{c}
	{\hat{\bf e}}_{i\parallel}\\
	{\hat{\bf e}}_{i\perp}
\end{array}
\right);
\label{eq:yz_vs_eis}
\end{equation}
substituting (\ref{eq:yz_vs_eis}) into (\ref{eq:fml_rel_S_v2}) we
obtain
\begin{equation}
\left(\!
\begin{array}{cc}
	S_2&S_3\\
	S_4&S_1
\end{array}
\!\right)
\left(\!
\begin{array}{c}
	{\bf E}_i(0)\cdot{\hat{\bf e}}_{i\parallel}\\
	{\bf E}_i(0)\cdot{\hat{\bf e}}_{i\perp}
\end{array}
\!\right)
=
i
\left(\!
\begin{array}{cc}
	-f_{11}&-f_{12}\\
	f_{21}&f_{22}
\end{array}
\!\right)
\left(\!
\begin{array}{cc}
	a&b\\
	c&d
\end{array}
\!\right)
\left(\!
\begin{array}{cc}
	\cos\phi_s&	\sin\phi_s\\
	\sin\phi_s&	-\cos\phi_s
\end{array}
\!\right)
\left(\!
\begin{array}{c}
	{\bf E}_i(0)\cdot{\hat{\bf e}}_{i\parallel}\\
	{\bf E}_i(0)\cdot{\hat{\bf e}}_{i\perp}
\end{array}
\!\right)
\label{eq:fml_rel_S_v3}
\end{equation}

Eq. (\ref{eq:fml_rel_S_v3}) must be true for all ${\bf E}_i(0)$, so we
obtain an expression for the complex scattering amplitude matrix in terms
of the $f_{ml}$:
\begin{equation}
\left(
\begin{array}{cc}
	S_2&S_3\\
	S_4&S_1
\end{array}
\right)
=
i
\left(
\begin{array}{cc}
	-f_{11}&-f_{12}\\
	f_{21}&f_{22}
\end{array}
\right)
\left(
\begin{array}{cc}
	a&	b\\
	c&	d
\end{array}
\right)
\left(
\begin{array}{cc}
	\cos\phi_s&\sin\phi_s\\
	\sin\phi_s&-\cos\phi_s
\end{array}
\right)~~~.
\label{eq:fml_rel_S_v4}
\end{equation}
This provides the 4 equations used in subroutine {\tt GETMUELLER} to
compute the amplitude scattering matrix elements:
\begin{eqnarray}
S_1 &=& -i\left[
	f_{21}(b\cos\phi_s-a\sin\phi_s)
	+f_{22}(d\cos\phi_s-c\sin\phi_s)
	\right]~~~,\label{eq:S_1_relto_f21andf22}\\
S_2 &=& -i\left[
	f_{11}(a\cos\phi_s+b\sin\phi_s)
	+f_{12}(c\cos\phi_s+d\sin\phi_s)
	\right]~~~,\\
S_3 &=& i\left[
	f_{11}(b\cos\phi_s-a\sin\phi_s)
	+f_{12}(d\cos\phi_s-c\sin\phi_s)
	\right]~~~,\\
S_4 &=& i\left[
	f_{21}(a\cos\phi_s+b\sin\phi_s)
	+f_{22}(c\cos\phi_s+d\sin\phi_s)
	\right] ~~~.
\end{eqnarray}

\subsection{Stokes Parameters}
It is both convenient and customary to characterize both incident and
scattered radiation by 4 ``Stokes parameters'' -- 
the elements of the ``Stokes vector''.
There are different conventions in the literature; we adhere to the
definitions of the Stokes vector ($I$,$Q$,$U$,$V$) adopted in the
excellent treatise by Bohren \&
Huffman (1983), to which the reader is referred for further detail.
Here are some examples of Stokes vectors $(I,Q,U,V)=(1,Q/I,U/I,V/I)I$:
\begin{itemize}
\item $(1,0,0,0)I$ : unpolarized light (with intensity $I$);
\item $(1,1,0,0)I$ : 100\% linearly polarized with ${\bf E}$ parallel to the
        scattering plane;
\item $(1,-1,0,0)I$ : 100\% linearly polarized with ${\bf E}$ perpendicular
        to the scattering plane;
\item $(1,0,1,0)I$ : 100\% linearly polarized with ${\bf E}$ at +45$^\circ$
        relative to the scattering plane;
\item $(1,0,-1,0)I$ : 100\% linearly polarized with ${\bf E}$ at -45$^\circ$
        relative to the scattering plane;
\item $(1,0,0,1)I$ : 100\% right circular polarization ({\it i.e.,} negative
        helicity);
\item $(1,0,0,-1)I$ : 100\% left circular polarization ({\it i.e.,} positive
        helicity).
\end{itemize}

\subsection{Relation Between Stokes Parameters of Incident and Scattered
Radiation: The Mueller Matrix}
It is convenient to describe the scattering properties in terms of
the $4\times4$ Mueller matrix relating the Stokes parameters 
$(I_i,Q_i,U_i,V_i)$ and
$(I_s,Q_s,U_s,V_s)$ of
the incident and scattered radiation:
\begin{equation}
\left(
\begin{array}{c}
	I_s\\
	Q_s\\
	U_s\\
	V_s
\end{array}
\right)
=
{1\over k^2r^2}
\left(
\begin{array}{cccc}
	S_{11}&S_{12}&S_{13}&S_{14}\\
	S_{21}&S_{22}&S_{23}&S_{24}\\
	S_{31}&S_{32}&S_{33}&S_{34}\\
	S_{41}&S_{42}&S_{43}&S_{44}
\end{array}
\right)
\left(
\begin{array}{c}
	I_i\\
	Q_i\\
	U_i\\
	V_i
\end{array}
\right)~~~.
\end{equation}
Once the amplitude scattering matrix elements are obtained, the Mueller
matrix elements can be computed (Bohren \& Huffman 1983):
\begin{eqnarray}
S_{11}&=&\left(|S_1|^2+|S_2|^2+|S_3|^2+|S_4|^2\right)/2~~~,\nonumber\\
S_{12}&=&\left(|S_2|^2-|S_1|^2+|S_4|^2-|S_3|^2\right)/2~~~,\nonumber\\
S_{13}&=&{\rm Re}\left(S_2S_3^*+S_1S_4^*\right)~~~,\nonumber\\
S_{14}&=&{\rm Im}\left(S_2S_3^*-S_1S_4^*\right)~~~,\nonumber\\
S_{21}&=&\left(|S_2|^2-|S_1|^2+|S_3|^2-|S_4|^2\right)/2~~~,\nonumber\\
S_{22}&=&\left(|S_1|^2+|S_2|^2-|S_3|^2-|S_4|^2\right)/2~~~,\nonumber\\
S_{23}&=&{\rm Re}\left(S_2S_3^*-S_1S_4^*\right)~~~,\nonumber\\
S_{24}&=&{\rm Im}\left(S_2S_3^*+S_1S_4^*\right)~~~,\nonumber\\
S_{31}&=&{\rm Re}\left(S_2S_4^*+S_1S_3^*\right)~~~,\nonumber\\
S_{32}&=&{\rm Re}\left(S_2S_4^*-S_1S_3^*\right)~~~,\nonumber\\
S_{33}&=&{\rm Re}\left(S_1S_2^*+S_3S_4^*\right)~~~,\nonumber\\
S_{34}&=&{\rm Im}\left(S_2S_1^*+S_4S_3^*\right)~~~,\nonumber\\
S_{41}&=&{\rm Im}\left(S_4S_2^*+S_1S_3^*\right)~~~,\nonumber\\
S_{42}&=&{\rm Im}\left(S_4S_2^*-S_1S_3^*\right)~~~,\nonumber\\
S_{43}&=&{\rm Im}\left(S_1S_2^*-S_3S_4^*\right)~~~,\nonumber\\
S_{44}&=&{\rm Re}\left(S_1S_2^*-S_3S_4^*\right)~~~.
\end{eqnarray}
These matrix elements are computed in {\tt DDSCAT} and passed to subroutine
{\tt WRITESCA} which handles output of scattering properties.
Although the Muller matrix has 16 elements, only 9 are independent.

The user can select up to 9 distinct Muller matrix elements to be
printed out in the output files 
{\tt w}{\it xx}{\tt r}{\it yy}{\tt k}{\it zzz}{\tt .sca} and
{\tt w}{\it xx}{\tt r}{\it yy}{\tt ori.avg}; this choice is made by 
providing a list of indices in {\tt ddscat.par} 
(see Appendix \ref{app:ddscat.par}).

If the user does not provide a list of elements, 
{\tt WRITESCA} will provide a ``default'' set of 6 selected elements: 
$S_{11}$, $S_{21}$, $S_{31}$, $S_{41}$ 
(these 4 elements describe the intensity and polarization 
state for scattering of unpolarized incident radiation), 
$S_{12}$, and $S_{13}$.

In addition, {\tt WRITESCA} writes out the linear polarization $P$ of the
scattered light for incident unpolarized light:
\begin{equation}
P = {(S_{21}^2+S_{31}^2)^{1/2}\over S_{11}} ~~~.
\end{equation}

\subsection{One Incident Polarization State Only ({\tt IORTH=1})}

In some cases it may be desirable to limit the calculations to a
single incident polarization state -- for example, when each solution
is very time-consuming, and the target is known to have some symmetry
so that solving for a single incident polarization state may be
sufficient for the required purpose.  In this case, set {\tt IORTH=1}
in {\tt ddscat.par}.

When {\tt IORTH=1}, only $f_{11}$ and $f_{21}$ are available;
hence, {{\bf DDSCAT}}\ cannot
automatically generate the Mueller matrix elements.
In this case, the output routine {\tt WRITESCA} writes out the quantities
$|f_{11}|^2$, $|f_{21}|^2$, ${\rm Re}(f_{11}f_{21}^*)$, and
${\rm Im}(f_{11}f_{21}^*)$ for each of the scattering directions.

Note, however, that if {\tt IPHI} is greater than 1, {{\bf DDSCAT}}\
will automatically set {\tt IORTH=2} even if {\tt ddscat.par}
specified {\tt IORTH=1}: this is because when more than one value of
the target orientation angle $\Phi$ is required, there is no
additional ``cost'' to solve the scattering problem for the second
incident polarization state, since when solutions are available for
two orthogonal states for some particular target orientation, the
solution may be obtained for another target orientation differing only
in the value of $\Phi$ by appropriate linear combinations of these
solutions.  Hence we may as well solve the ``complete'' scattering
problem so that we can compute the complete Mueller matrix.

\section{Using MPI (Message Passing Interface)
	\label{sec:MPI}}

Many (probably most) 
users of {\bf DDSCAT} will wish to calculate scattering and absorption
by a given target for more than one orientation relative to the
incident radiation (e.g., when calculating orientational averages). 
{\tt DDSCAT} can automatically
carry out the calculations over different target orientations.  Normally,
these calculations will be carried out sequentially.
However, on a multiprocessor system [either a symmetric multiprocessor
(SMP) system, or a cluster of networked computers]
with {\tt MPI} (Message Passing Interface) software installed,
\ddscat\ allows the user to carry out
parallel calculations of scattering by different target orientations,
with the results gathered together and with orientational averages
calculated just as they are when sequential calculations are carried out.
Note that 
the MPI-capable executable can also be used for ordinary serial calculations
using a single cpu.

More than one implementation of {\tt MPI} exists.  
{\tt MPI} support within \ddscat\ is compliant with MPI-1.2 and MPI-2
standards\footnote{\tt http://www.mpi-forum.org/}, 
and should be usable under any implementation of {\tt MPI}
that is compatible with those standards.

At Princeton University
Department of Astrophysical Sciences we are using 
{\tt MPICH}\footnote{\tt http://www.mcs.anl.gov/mpi/mpich },
a publicly-available implementation of {\tt MPI}.

\subsection{Source Code for the MPI-compatible version of DDSCAT}

In order to use DDSCAT with MPI, the user must first make sure that
the correct version of the code is being used.
To prepare the MPI-capable version:
\begin{enumerate}
\item In {\tt DDSCAT.f} :
	\begin{itemize}
	\item Make sure the line\\
		\hspace*{4em}{\tt INCLUDE 'mpif.h'}\\
		is not commented out.
	\item Make sure the line\\
		{\tt C\hspace*{3em}INTEGER MPI\_COMM\_WORLD}\\
		{\it is} commented out.
	\end{itemize}
\item In {\tt mpisubs.f, SUBROUTINE COLSUM} :
	\begin{itemize}
	\item Make sure the line\\
		\hspace*{4em}{\tt INCLUDE 'mpif.h'}\\
		is not commented out.
	\item Make sure the line\\
		{\tt C\hspace*{3em}INTEGER MPI\_COMM\_WORLD,MPI\_SUM,MPI\_REAL,MPI\_COMPLEX}\\
		{\it is} commented out.
	\end{itemize}
\item In {\tt mpisubs.f, SUBROUTINE SHARE1} :
	\begin{itemize}
	\item Make sure the line\\
		\hspace*{4em}{\tt INCLUDE 'mpif.h'}\\
		is not commented out.
	\item Make sure the lines\\
		{\tt C\hspace*{3em}INTEGER MPI\_CHARACTER,MPI\_COMM\_WORLD,MPI\_INTEGER,\\
		C\hspace*{2.5em}\&\hspace*{2em}MPI\_INTEGER2,MPI\_REAL,MPI\_COMPLEX}
	
		{\it are} commented out.
	\end{itemize}
\item In {\tt mpisubs.f, SUBROUTINE SHARE2} :
	\begin{itemize}
	\item Make sure the line\\
		\hspace*{4em}{\tt INCLUDE 'mpif.h'}\\
		is not commented out.
	\item Make sure the line\\
		{\tt C\hspace*{3em}INTEGER MPI\_COMM\_WORLD,MPI\_REAL,MPI\_COMPLEX}
	
		{\it is} commented out.
	\end{itemize}
\end{enumerate}

\subsection{\label{subsec:compiling_mpi}
	Compiling and Linking the MPI-compatible version of DDSCAT}
In the {\tt Makefile}, enable the following definitions:\\
{\tt
\hspace*{3em}MPIsrc\hspace*{3em}= mpi\_subs.f\\
\hspace*{3em}MPIobj\hspace*{3em}= mpi\_subs.o}

\noindent At Princeton University Dept.\ of Astrophysical Sciences we are 
currently using
{\tt mpich-1.2.4}\footnote{\tt http://www-unix.mcs.anl.gov/mpi/mpich/}.
{\tt mpich} provides a ``compiler wrapper'' {\tt mpif77} that is supposed to
automatically set the variables {\tt MPICH\_F77} and {\tt MPICH\_F77LINKER}
to values appropriate for use of the {\tt pgf77} compiler from
PGI\footnote{\tt http://www.pgroup.com} which is installed on our
beowulf cluster.  Other compilers, e.g., the Intel f77 compiler, can
also be employed.

\subsection{Code Execution Under MPI}

Local installations of {\tt MPI} will vary -- you should consult with
someone familiar with way {\tt MPI} is installed and used on your system.

At Princeton University Dept.\ of Astrophysical Sciences we use {\tt PBS}
(Portable Batch System)\footnote{\tt http://www.openpbs.org}
to schedule jobs.
{\tt MPI} jobs are submitted using {\tt PBS} by first creating a 
shell script such as the following example file {\tt pbs.submit}:
\begin{verbatim}
     #!/bin/bash
     #PBS -l nodes=2:ppn=1
     #PBS -l mem=1200MB,pmem=300MB
     #PBS -m bea
     #PBS -j oe
     cd $PBS_O_WORKDIR
     /usr/local/bin/mpiexec ddscat
\end{verbatim}
The lines beginning with {\tt\#PBS -l} specify the required resources:\\
{\tt\#PBS -l nodes=2:ppn=1} specifies that 2 nodes are to be used,
with 1 processor per node.\\
{\tt\#PBS -l mem=1200MB,pmem=300MB} specifies that the total memory
required ({\tt mem}) is 1200MB, 
and the maximum physical memory used by any single
process ({\tt pmem}) is 300MB.  The actual definition of {\tt mem} is
not clear, but in practice it seems that it should be set equal to
2$\times$({\tt nodes})$\times$({\tt ppn})$\times$({\tt pmem}) .\\
{\tt\#PBS -m bea} specifies that PBS should send email when the job begins
({\tt b}), and when it ends ({\tt e}) or aborts ({\tt a}).\\
{\tt\#PBS -j oe} specifies that the output from stdout and stderr will be
merged, intermixed, as stdout.

\noindent This example assumes that the executable {\tt dscat} is located
in the same directory where the code is to execute and write its output.
If {\tt ddscat} is located in another directory, simply give the full pathname.
to it.
The {\tt qsub} command is used to submit the {\tt PBS} job:
\begin{verbatim}
% qsub pbs.submit
\end{verbatim}

As the calculation proceeds, the usual 
output files will be written to this directory: for each wavelength, target size,
and target orientation,
there will be a file
{\tt w}{\it aa}{\tt r}{\it bb}{\tt k}{\it ccc}{\tt .sca}, where
{\it aa=00, 01, 02, ...} specifies the wavelength,
{\it bb=00, 01, 02, ...} specifies the target size,
and
{\it cc=000, 001, 002, ...} specifies the orientation.
For each wavelength and target size there will also be a file
{\tt w}{\it aa}{\tt r}{\it bb}{\tt ori.avg} with orientationally-averaged
quantities.
Finally, there will also be tables {\tt qtable},
and {\tt qtable2} with orientationally-averaged cross sections for each
wavelength and target size.

In addition, each processor employed will write to its own log file
{\tt ddscat.log\_}{\it nnn}, where {\it nnn=000, 001, 002, ...}.
These files contain information
concerning cpu time consumed by different parts of the calculation,
convergence to the specified error tolerance, etc.  If you are uncertain
about how the calculation proceeded, examination of these log files
is recommended.

\section{Graphics and Postprocessing\label{sec:postprocessing}}

\subsection{IDL}

At present, we do not offer a comprehensive  package for {{\bf DDSCAT}}\
data postprocessing and graphical display in IDL. However, there are several
developments  worth mentioning:
First, we offer several output capabilities from within {{\bf DDSCAT}}:
ASCII (see \S\ref{subsec:ascii}), 
FORTRAN unformatted binary (see \S\ref{subsec:binary}), 
and netCDF portable binary (see \S\ref{subsec:netCDF}).
Second, we offer several skeleton IDL utilities:

\begin{itemize}
\item {\tt bhmie.pro} is our translation to IDL of the popular Bohren-Huffman
code which calculates efficiencies for spherical particles using Mie theory.

\item {\tt readbin.pro} reads the FORTRAN unformatted binary file written by
routine {\tt writebin.f}. The variables are stored in a {\tt common} block.

\item {\tt readnet.pro} reads NetCDF portable binary file and should be the
method of choice for IDL users. It offers random data access.

\item {\tt mie.pro} is an example of an interface to binary files and
the Bohren-Huffman code.  It plots a comparison of {{\bf DDSCAT}}\ results 
with scattering by equivalent radius spheres.

\end{itemize}
\noindent At present the IDL code is experimental.

\section{Miscellanea}

Additional source code, refractive index files, etc., contributed by users
will be located in the directory {\tt DDA/misc}.  
These routines and files should be
considered to be {\bf not supported} by Draine and Flatau -- 
{\it caveat receptor!}  
These routines and files should be accompanied by enough information (e.g.,
comments in source code) to explain their use.

\section{Finale}

This User Guide is somewhat inelegant, but we hope that it will prove
useful.  The structure of the {\tt ddscat.par} file is intended to be
simple and suggestive so that, after reading the above notes once, the
user may not have to refer to them again.

The file {\tt rel\_notes} in {\tt DDA/doc} lists known bugs in 
\ddscat\ .  
Up-to-date release notes will be available at the {\bf DDSCAT} web site,\\
\hspace*{2em}{\tt http://www.astro.princeton.edu/}$\sim${\tt draine/DDSCAT.html}

Users are encouraged to provide B. T. Draine 
({\tt draine@astro.princeton.edu}) with their
email address; bug reports, and any new releases of {{\bf DDSCAT}}, 
will be made known to those who do!

P. J. Flatau maintains the WWW page 
``SCATTERLIB - Light Scattering Codes Library"
with URL {\tt http://atol.ucsd.edu/$\sim$pflatau}.
The SCATTERLIB Internet site is a library of light scattering codes. 
Emphasis is on providing source
codes (mostly FORTRAN). However, other information related to scattering 
on spherical and
non-spherical particles is collected: an extensive list of references 
to light scattering methods, refractive index, etc. 
This URL page contains section on the discrete dipole approximation.

\bigskip
Concrete suggestions for improving {{\bf DDSCAT}}\ (and this User Guide) are
welcomed.
If you wish to cite this User Guide, we suggest the following 
citation:\\

Draine, B.T., \& Flatau, P.J. 2000, 
``User Guide for the Discrete
Dipole Approximation Code DDSCAT (Version 6.0)'',
http://arxiv.org/abs/astro-ph/??????

\bigskip
Finally, the authors have one special request:
We would very much appreciate preprints and (especially!) 
reprints of any papers which make use of {{\bf DDSCAT}}!

\section{Acknowledgments}

\begin{itemize}
\item The routine {\tt ESELF} making use of the FFT was originally written by 
Jeremy Goodman, Princeton University Observatory.  
\item The FFT routine {\tt FOURX} (no longer used by \ddscat, but
included for comparison of FFT routines by {\tt TSTFFT}) 
is based on a FFT routine written by Norman
Brenner (Brenner 1969). 
\item The routine {\tt REFICE} was written
by Steven B. Warren, based on Warren (1984). 
\item The routine {\tt REFWAT} was written by Eric A. Smith.  
\item The {\tt GPFAPACK} package was written by Clive Temperton (1992), and
generously made available by him for use with {\tt DDSCAT}.
\item We make use of routines from the 
{\tt LAPACK}\footnote{\tt http://netlib.org} 
package (Anderson {\it et al.} 1995),
the result of work by Jack Dongarra and others at the Univ. of Tennessee,
Univ. of California Berkeley, NAG Ltd., Courant Institute, Argonne National
Lab, and Rice University.
\item The {\tt FFTW}\footnote{\tt http://www.fftw.org} package, 
to which we link, was
written by Matteo Frigo and Steven G. Johnson ({\tt fftw@fftw.org}).
\item Much of the work involved in modifying {\tt DDSCAT} to use {\tt MPI} 
was done by Matthew Collinge, Princeton University.
\end{itemize}
We are deeply 
indebted to all of these authors for making their code available.  

We wish also to acknowledge bug reports and suggestions from {{\bf DDSCAT}}
users, including
Henrietta Lemke, Timo Nousianen, and Mike Wolff.

Development of {{\bf DDSCAT}}\ was supported in part by National Science Foundation
grants AST-8341412, AST-8612013, AST-9017082, AST-9319283, 
AST-9616429, AST-9988126 to BTD,
in part by support from the Office of Naval Research 
Young Investigator Program to PJF, in part by DuPont Corporate Educational
Assistance to PJF, and in part by the United Kingdom Defence Research Agency.

\vfill\eject
\appendix
\section{Understanding and Modifying {\tt ddscat.par}\label{app:ddscat.par}}

In order to use DDSCAT to perform the specific calculations of interest to
you, it will be necessary to modify the {\tt ddscat.par} file.  
Here we list the sample {\tt ddscat.par} file, followed by a discussion of
how to modify this file as needed.
Note that all numerical input data in DDSCAT is read with free-format 
{\tt READ(IDEV,*)...} statements.  
Therefore you do not need to worry about the precise format in which integer 
or floating point numbers are entered on a line.
The crucial thing is that lines in {\tt ddscat.par} containing numerical 
data have the correct number of data entries, with any informational 
comments appearing {\it after} the numerical data on a given line.
{\footnotesize
\begin{verbatim}
' =================== Parameter file ===================' 
'**** PRELIMINARIES ****'
'NOTORQ'= CMTORQ*6 (DOTORQ, NOTORQ) -- either do or skip torque calculations
'PBCGST'= CMDSOL*6 (PBCGST, PETRKP) -- select solution method
'GPFAFT'= CMETHD*6 (GPFAFT, FFTW21, CONVEX)
'LATTDR'= CALPHA*6 (LATTDR is only supported option now)
'NOTBIN'= CBINFLAG (ALLBIN, ORIBIN, NOTBIN)
'NOTCDF'= CNETFLAG (ALLCDF, ORICDF, NOTCDF)
'RCTNGL'= CSHAPE*6 (FRMFIL,ELLIPS,CYLNDR,RCTNGL,HEXGON,TETRAH,UNICYL,UNIELL)
8. 6. 4. = shape parameters PAR1, PAR2, PAR3
1         = NCOMP = number of dielectric materials
'TABLES'= CDIEL*6 (TABLES,H2OICE,H2OLIQ; if TABLES, then filenames follow...)
'diel.tab' = name of file containing dielectric function
'**** CONJUGATE GRADIENT DEFINITIONS ****'
0       = INIT (TO BEGIN WITH |X0> = 0)
1.00e-5 = ERR = MAX ALLOWED (NORM OF |G>=AC|E>-ACA|X>)/(NORM OF AC|E>)
'**** ANGLES FOR CALCULATION OF CSCA, G'
33      = ICTHM (number of theta values for evaluation of Csca and g)
12      = IPHM (number of phi values for evaluation of Csca and g)
'**** Wavelengths (micron) ****'
6.283185 6.283185 1 'INV' = wavelengths (first,last,how many,how=LIN,INV,LOG)
'**** Effective Radii (micron) **** '
1. 1. 1  'LIN' = eff. radii (first, last, how many, how=LIN,INV,LOG)
'**** Define Incident Polarizations ****'
(0,0) (1.,0.) (0.,0.) = Polarization state e01 (k along x axis)
2 = IORTH  (=1 to do only pol. state e01; =2 to also do orth. pol. state)
1 = IWRKSC (=0 to suppress, =1 to write ".sca" file for each target orient.
'**** Prescribe Target Rotations ****'
 0.   0.  1  = BETAMI, BETAMX, NBETA (beta=rotation around a1)
 0.  90.  3  = THETMI, THETMX, NTHETA (theta=angle between a1 and k)
 0.   0.  1  = PHIMIN, PHIMAX, NPHI (phi=rotation angle of a1 around k)
'**** Specify first IWAV, IRAD, IORI (normally 0 0 0) ****'
0   0   0    = first IWAV, first IRAD, first IORI (0 0 0 to begin fresh)
'**** Select Elements of S_ij Matrix to Print ****'
6	= NSMELTS = number of elements of S_ij to print (not more than 9)
11 12 21 22 31 41	= indices ij of elements to print
'**** Specify Scattered Directions ****'
0.  0. 180. 30 = phi, thetan_min, thetan_max, dtheta (in degrees) for plane A
90. 0. 180. 30 = phi, ... for plane B
\end{verbatim}
}
\begin{tabular}{l l}
Lines	&Comments\\
1-2	&comment lines -- no need to change.\\
3	&{\tt NOTORQ} if torque calculation is not required; \\
	&{\tt DOTORQ} if torque calculation is required. \\
4	&{\tt PBCGST} is recommended; 
	other option is {\tt PETRKP} (see \S\ref{sec:choice_of_algorithm}).\\
5	&{\tt GPFAFT} is supplied as default, but {\tt FFTW21} is recommended 
	if {\tt DDSCAT} has been compiled with\\
	&{\tt FFTW} support (see \S\S\ref{subsec:support_for_FFTW},
	\protect{\ref{sec:choice_of_fft}}); only other option is
	{\tt CONVEX} (valid only on a Convex computer).\\
6	&{\tt LATTDR} (Draine \& Goodman [2000] 
	Lattice Dispersion Relation is only option now supported)\\
	&(see \S\protect{\ref{sec:polarizabilities}}).\\
7	&{\tt ALLBIN} for full unformatted binary dump (\S\ref{subsec:binary}); \\
	&{\tt ORIBIN} for unformatted binary dump of orientational averages only; \\
	&{\tt NOTBIN} for no unformatted binary output.\\
8	&{\tt ALLCDF} for output in netCDF format (must have netCDF option enabled; cf. \S\ref{subsec:netCDF}); \\
	&{\tt ORICDF} for orientational averages in netCDF format (must have netCDF option enabled).\\
	&{\tt NOTCDF} for no output in netCDF format; \\
9	&specify choice of target shape (see \S\ref{sec:target_generation} for
	description of options {\tt RCTNGL}, {\tt ELLIPS}, {\tt TETRAH}, ...)\\
10	&shape parameters {SHPAR1}, {SHPAR2}, {SHPAR3}, ... 
	(see \S\ref{sec:target_generation}).\\
11	&number of different dielectric constant tables (\S\ref{sec:dielectric_func}).\\
12	&{\tt TABLES} -- dielectric function to be provided via input table.\\
13	&name(s) of dielectric constant table(s) (one per line).\\
14	&comment line -- no need to change.\\
15	&{\tt 0} is recommended value of parameter {\tt INIT}.\\
16	&{\tt ERR} = error tolerance $h$: maximum allowed value of 
	$|A^\dagger E-A^\dagger AP|/|A^\dagger E|$ [see eq.(\ref{eq:err_tol})].\\
17	&comment line -- no need to change.\\
18	&{\tt ICTHM} -- number of $\theta_s$ values for angular averages (\S\ref{sec:averaging_scattering}).\\
19	&{\tt IPHM} -- number of $\phi_s$ values for angular averages.\\
20	&comment line -- no need to change.\\
21	&$\lambda$ -- first, last, how many, how chosen.\\
22	&comment line -- no need to change.\\
23	&$a_{\rm eff}$ -- first, last, how many, how chosen.\\
24	&comment line -- no need to change.\\
25	&specify x,y,z components of (complex) incident polarization ${\hat{\bf e}}_{01}$ (\S\ref{sec:incident_polarization})\\
26	&{\tt IORTH} = 1 to do one polarization state only;\\
	&2 to do second (orthogonal) incident polarization as well.\\
27	&{\tt IWRKSC} = 0 to suppress writing of ``.sca" files;\\
	&2 to enable writing of ``.sca" files.\\
28	&comment line -- no need to change.\\
29	&$\beta$ (see \S\ref{sec:target_orientation}) -- first, last, how many .\\
30	&$\Theta$ -- first, last, how many.\\
31	&$\Phi$ -- first, last, how many.\\
32	&comment line -- no need to change.\\
33	&{\tt IWAV0 IRAD0 IORI0} -- specify starting values of integers
	{\tt IWAV IRAD IORI} (normally {\tt 0 0 0}).\\
34	&comment line -- no need to change.\\
35	&$N_{S}$ = number of scattering matrix elements (must be $\leq9$)\\
36	&indices $ij$ of $N_S$ elements of the scattering matrix $S_{ij}$\\
37	&comment line -- no need to change.\\
38	&$\phi_s$ for first scattering plane, $\theta_{s,min}$, $\theta_{s,max}$, how many $\theta_s$ values;\\
39,...	&$\phi_s$ for 2nd,... scattering plane, ...
\end{tabular}

\section{{\tt w{\it xx}r{\it yy}ori.avg} Files\label{app:w00r00ori.avg}}

{\small
The file {\tt w00r00ori.avg} contains the results for the first wavelength
({\tt w00}) and first target radius ({\tt r00})
averaged over orientations ({\tt ori.avg}).
The {\tt w00r00ori.avg} file generated by the sample calculation should look 
like the following:
\vspace*{-0.4em}
{\footnotesize
\begin{verbatim}
 DDSCAT --- DDSCAT.6.0 [03.07.13] 
 TARGET --- Rectangular prism; NX,NY,NZ=   8   6   4                          
 GPFAFT --- method of solution 
 LATTDR --- prescription for polarizabilies
 RCTNGL --- shape 
    192 = NAT0 = number of dipoles

  AEFF=   1.00000 = effective radius (physical units)
  WAVE=   6.28319 = wavelength (physical units)
K*AEFF=   1.00000 = 2*pi*aeff/lambda
n= ( 1.3300 ,  0.0100),  eps.= (  1.7688 ,  0.0266)  |m|kd=  0.3716 for subs. 1
   TOL= 1.000E-05 = error tolerance for CCG method
 ICTHM= 33 = theta values used in comp. of Qsca,g
 IPHIM= 12 = phi values used in comp. of Qsca,g
 ( 1.00000  0.00000  0.00000) = target axis A1 in Target Frame
 ( 0.00000  1.00000  0.00000) = target axis A2 in Target Frame
 ( 0.27942  0.00000  0.00000) = k vector (latt. units) in Lab Frame
 ( 0.00000, 0.00000)( 1.00000, 0.00000)( 0.00000, 0.00000)=inc.pol.vec. 1 in LF
 ( 0.00000, 0.00000)( 0.00000, 0.00000)( 1.00000, 0.00000)=inc.pol.vec. 2 in LF
   0.000   0.000 = beta_min, beta_max ;  NBETA = 1
   0.000  90.000 = theta_min, theta_max; NTHETA= 3
   0.000   0.000 = phi_min, phi_max   ;   NPHI = 1
 Results averaged over   3 target orientations
                   and   2 incident polarizations
          Qext       Qabs        Qsca     g(1)=<cos>     Qbk        Qpha
 JO=1:  1.3296E-01 3.2687E-02 1.0028E-01  2.3451E-01 5.5520E-03 4.8356E-01
 JO=2:  9.0628E-02 2.3972E-02 6.6655E-02  2.6726E-01 3.4303E-03 4.1588E-01
 mean:  1.1180E-01 2.8330E-02 8.3466E-02  2.4758E-01 4.4912E-03 4.4972E-01
 Qpol=  4.2337E-02                                       dQpha= 6.7685E-02
         Qsca*g(1)   Qsca*g(2)   Qsca*g(3)   iter  mxiter
 JO=1:  2.3516E-02  1.1256E-03  3.2952E-10      3    576
 JO=2:  1.7814E-02  2.7678E-03  7.6056E-10      3    576
 mean:  2.0665E-02  1.9467E-03  5.4504E-10
            ** Mueller matrix elements for selected scattering directions **
theta   phi    Pol.    S_11       S_12       S_21       S_22       S_31       S_41
  0.0   0.0  0.15320  5.167E-02  7.916E-03  7.916E-03  5.167E-02  8.165E-11  2.565E-11
 30.0   0.0  0.01117  4.046E-02  4.518E-04  4.518E-04  4.046E-02  7.106E-12  2.170E-11
 60.0   0.0  0.47184  2.169E-02 -1.023E-02 -1.023E-02  2.169E-02  3.183E-11 -5.271E-11
 90.0   0.0  0.99888  1.193E-02 -1.192E-02 -1.192E-02  1.193E-02  5.394E-12  5.443E-11
120.0   0.0  0.48743  1.170E-02 -5.702E-03 -5.702E-03  1.170E-02 -3.811E-11  3.570E-11
150.0   0.0  0.06956  1.403E-02  9.758E-04  9.758E-04  1.403E-02 -4.999E-11 -1.284E-11
180.0   0.0  0.23621  1.411E-02  3.333E-03  3.333E-03  1.411E-02 -1.220E-11 -4.503E-12
  0.0  90.0  0.15320  5.167E-02 -7.916E-03 -7.916E-03  5.167E-02 -7.224E-10  2.865E-11
 30.0  90.0  0.28550  4.487E-02 -1.281E-02 -1.280E-02  4.486E-02 -5.036E-04  8.979E-05
 60.0  90.0  0.67857  3.040E-02 -2.063E-02 -2.062E-02  3.039E-02 -7.990E-04  1.643E-04
 90.0  90.0  0.99926  1.991E-02 -1.991E-02 -1.989E-02  1.989E-02 -7.361E-04  1.827E-04
120.0  90.0  0.72285  1.626E-02 -1.176E-02 -1.175E-02  1.625E-02 -4.415E-04  1.359E-04
150.0  90.0  0.35833  1.478E-02 -5.295E-03 -5.293E-03  1.478E-02 -1.716E-04  6.554E-05
180.0  90.0  0.23621  1.411E-02 -3.333E-03 -3.333E-03  1.411E-02  3.323E-10 -2.245E-11
\end{verbatim}
}
}
\section{{\tt w{\it xx}r{\it yy}k{\it zzz}.sca} Files
	\label{app:w00r00k000.sca}}

The {\tt w00r00k000.sca} file contains the results for the first
wavelength ({\tt w00}), first target radius ({\tt r00}),
and first orientation ({\tt k000}).
The {\tt w00r00k000.sca} file created by the sample calculation should look
like the following:
{\footnotesize
\begin{verbatim}
 DDSCAT --- DDSCAT.6.0 [03.07.13] 
 TARGET --- Rectangular prism; NX,NY,NZ=   8   6   4                          
 GPFAFT --- method of solution 
 LATTDR --- prescription for polarizabilies
 RCTNGL --- shape 
    192 = NAT0 = number of dipoles

  AEFF=   1.00000 = effective radius (physical units)
  WAVE=   6.28319 = wavelength (physical units)
K*AEFF=   1.00000 = 2*pi*aeff/lambda
n= ( 1.3300 ,  0.0100),  eps.= (  1.7688 ,  0.0266)  |m|kd=  0.3716 for subs. 1
   TOL= 1.000E-05 = error tolerance for CCG method
 ICTHM= 33 = theta values used in comp. of Qsca,g
 IPHIM= 12 = phi values used in comp. of Qsca,g
 ( 1.00000  0.00000  0.00000) = target axis A1 in Target Frame
 ( 0.00000  1.00000  0.00000) = target axis A2 in Target Frame
 ( 0.27942  0.00000  0.00000) = k vector (latt. units) in TF
 ( 0.00000, 0.00000)( 1.00000, 0.00000)( 0.00000, 0.00000)=inc.pol.vec. 1 in TF
 ( 0.00000, 0.00000)( 0.00000, 0.00000)( 1.00000, 0.00000)=inc.pol.vec. 2 in TF
 BETA =  0.000 = rotation of target around A1
 THETA=  0.000 = angle between A1 and k
  PHI =  0.000 = rotation of A1 around k
          Qext       Qabs        Qsca     g(1)=<cos>     Qbk        Qpha
 JO=1:  1.1104E-01 3.0279E-02 8.0766E-02  3.5035E-01 1.8580E-03 4.6680E-01
 JO=2:  8.6508E-02 2.4413E-02 6.2095E-02  3.6498E-01 1.3620E-03 4.1968E-01
 mean:  9.8776E-02 2.7346E-02 7.1430E-02  3.5671E-01 1.6100E-03 4.4324E-01
 Qpol=  2.4537E-02                                       dQpha= 4.7115E-02
         Qsca*g(1)   Qsca*g(2)   Qsca*g(3)   iter  mxiter
 JO=1:  2.8296E-02 -5.5908E-10 -1.2161E-09      3    576
 JO=2:  2.2663E-02  3.0799E-09  1.4463E-09      3    576
 mean:  2.5480E-02  1.2604E-09  1.1509E-10
            ** Mueller matrix elements for selected scattering directions **
theta   phi    Pol.    S_11       S_12       S_21       S_22       S_31       S_41
  0.0   0.0  0.10772  8.312E-03  8.954E-04  8.954E-04  8.312E-03 -1.837E-11  5.826E-12
 30.0   0.0  0.02296  6.734E-03 -1.546E-04 -1.546E-04  6.734E-03  7.358E-13  9.139E-12
 60.0   0.0  0.48383  3.651E-03 -1.766E-03 -1.766E-03  3.651E-03 -5.798E-12 -1.051E-11
 90.0   0.0  0.99794  1.764E-03 -1.760E-03 -1.760E-03  1.764E-03 -6.232E-12  7.791E-12
120.0   0.0  0.53222  1.251E-03 -6.658E-04 -6.658E-04  1.251E-03 -1.813E-12 -4.691E-13
150.0   0.0  0.00167  9.868E-04 -1.644E-06 -1.644E-06  9.868E-04 -5.853E-12 -2.595E-12
180.0   0.0  0.15403  8.430E-04  1.298E-04  1.298E-04  8.430E-04  3.319E-13 -3.020E-12
  0.0  90.0  0.10772  8.312E-03 -8.954E-04 -8.954E-04  8.312E-03 -7.445E-11  9.616E-12
 30.0  90.0  0.24082  7.141E-03 -1.720E-03 -1.720E-03  7.141E-03 -7.673E-11 -6.021E-12
 60.0  90.0  0.64798  4.560E-03 -2.955E-03 -2.955E-03  4.560E-03 -2.574E-11  3.576E-12
 90.0  90.0  0.99942  2.566E-03 -2.565E-03 -2.565E-03  2.566E-03 -5.111E-12 -5.125E-12
120.0  90.0  0.68918  1.631E-03 -1.124E-03 -1.124E-03  1.631E-03  8.146E-12 -8.819E-13
150.0  90.0  0.28491  1.065E-03 -3.034E-04 -3.034E-04  1.065E-03  9.598E-12  1.278E-12
180.0  90.0  0.15403  8.430E-04 -1.298E-04 -1.298E-04  8.430E-04  7.621E-12 -1.666E-12
\end{verbatim}
}
\end{document}